\documentclass[preprint,aps,showpacs,nofootinbib,showkeys,
byrevtex]{revtex4}
\usepackage{amsmath}
\usepackage{graphicx}
\usepackage{latexsym}
\usepackage[normalem]{ulem}
\def\){\right)} 
\def\({\left(} 
\def\]{\right]} 
\def\[{\left[}

\def\chpt{\raise0.4ex\hbox{$\chi$}PT}
\def\schpt{S\raise0.4ex\hbox{$\chi$}PT}

\def\psibar{\overline{\psi}}

\newcommand{\eqn}[1]{\label{eq:#1}}
\newcommand{\refeq}[1]{(\ref{eq:#1})} 
\newcommand{\eq}{eq.~\refeq}
\newcommand{\eqs}[2]{eqs.~(\ref{eq:#1}-\ref{eq:#2})}

\newcommand{\Equation}{Equation~\refeq} 

\def\figref#1{Fig.~\ref{fig:#1}}

\def\eqstwo#1#2{Eqs.~(\ref{eq:#1}) and (\ref{eq:#2})}
\def\eqsthree#1#2#3{Eqs.~(\ref{eq:#1}), (\ref{eq:#2}) and (\ref{eq:#3})}
\def\eqsfour#1#2#3#4{Eqs.~(\ref{eq:#1}), (\ref{eq:#2}), (\ref{eq:#3}) and (\ref{eq:#4})}

\def\Equations#1#2{Equations~(\ref{eq:#1}) and (\ref{eq:#2})}

\newcommand{\mcal}[1]{{\mathcal #1}}

\newcommand{\be}{\begin{equation}}
\newcommand{\ee}{\end{equation}}
\newcommand{\ba}{\begin{array}}
\newcommand{\ea}{\end{array}}
\newcommand{\bea}{\begin{eqnarray}}
\newcommand{\eea}{\end{eqnarray}}
\newcommand{\beq}{\begin{eqnarray}}
\newcommand{\eeq}{\end{eqnarray}}
\newcommand{\pref}[1]{(\ref{#1})}
\newcommand{\nn}{\nonumber}

\def\cC{{\cal C}}

\def\cF{{\cal F}}

\def\cM{{\cal M}}

\def\cO{{\rm O}}

\def\cU{{\cal U}}

\def\ie{{\it i.e.},\ }
\def\eg{{\it e.g.},\ }

\begin{document}

\preprint{UTHEP-503, LA-UR-04-8892, BUHEP-05-02}

\title{Chiral Perturbation Theory for Staggered Sea Quarks and Ginsparg-Wilson Valence Quarks}
\author{Oliver B\"ar~${}^{a}$\footnote{Email: {\tt obaer@het.ph.tsukuba.ac.jp}},
Claude Bernard~${}^{b}$\footnote{Email: {\tt cb@lump.wustl.edu}} 
Gautam Rupak~${}^{c}$\footnote{Email: {\tt grupak@lanl.gov}} 
and
Noam Shoresh~${}^{d}$\footnote{Email: {\tt shoresh@cgr.harvard.edu}}
}
\affiliation{${}^a$ Institute of Physics, University of Tsukuba, Tsukuba, Ibaraki 305-8571, Japan.\\
${}^b$Department of Physics, Washington University, St.~Louis, MO 63130, USA.\\
${}^c$Los Alamos National Laboratory, Theoretical Division T-16, 
Los Alamos, NM 87545, USA. \\
${}^d$Department of Physics, Boston University, Boston, MA 02215, USA.}

%
\begin{abstract}
We study lattice QCD with staggered sea and Ginsparg-Wilson valence quarks. The Symanzik effective action for this mixed lattice theory, 
including the lattice spacing contributions of $\mcal O(a^{2})$,
is derived. Using this effective theory we construct the leading order chiral Lagrangian. The masses and decay constants of pseudoscalars containing two Ginsparg-Wilson valence quarks are computed at one loop order. 
\end{abstract}

\maketitle 

\section{Introduction}
\label{introduction}
%

Unquenched simulations with fermions that satisfy the Ginsparg-Wilson relation \cite{Ginsparg:1982bj} 
are computationally much more demanding than those with staggered or Wilson fermions. A precise comparison between the numerical cost depends on many details of the simulation, for example the lattice spacing, the quark masses and also the practical implementation of the Ginsparg-Wilson (GW)
fermion, 
\ie how well the overlap \cite{Narayanan:1992wx,Narayanan:1993sk,Narayanan:1994gw} or the domain-wall fermion \cite{Kaplan:1992bt,Furman:1994ky} is approximated. A recent review \cite{Kennedy:2004ae}, however, suggests that dynamical Ginsparg-Wilson fermions 
may be about ten to hundred times more expensive than corresponding simulations at comparable masses
with either improved 
staggered fermions using the Asqtad action \cite{Bernard:2001av,Aubin:2004wf,Aubin:2004fs} or twisted-mass Wilson fermions \cite{Frezzotti:2000nk}. For that reason most dynamical GW simulations so far have been carried out on small volumes together with rather heavy quark masses (see Ref.\ \cite{Kennedy:2004ae} and references therein). Simulating volumes and quark masses comparable to those in present-day staggered simulations \cite{Aubin:2004fs}, for example, is out of reach in the near future. 

A computationally affordable compromise for certain applications may be so-called mixed fermion  simulations. This type of simulation employs GW fermions only for the valence quarks, while the sea quarks are either staggered or Wilson fermions. In such an approach at least the valence sector exhibits all the benefits stemming from the exact chiral symmetry at non-zero lattice spacing \cite{Luscher:1998pq}. Moreover, provided one can use already existing unquenched configurations generated with either staggered or Wilson fermions, mixed simulations only require the computation of correlators in the background of these configurations. The additional numerical cost is therefore comparable to quenched GW fermion simulations. Such mixed actions in two dimensions were studied in the
Schwinger model in Ref.~\cite{Durr:2003xs}.  In four dimensions,
 preliminary results using the publicly available MILC configurations together with domain-wall or overlap valence fermions have been reported recently \cite{Renner:2004ck,Bowler:2004kv,Bowler:2004hs}. 

In this paper we construct the low-energy chiral effective Lagrangian for a mixed lattice theory with staggered sea and GW valence quarks. Based on this effective Lagrangian we compute the pseudoscalar meson masses and decay constants to one loop. Our results provide the leading quark mass and lattice spacing dependence of these quantities. The formulae are needed in order to analyze numerical data of mixed simulations.

This paper parallels Refs.\ \cite{Bar:2002nr,Bar:2003mh}, where the chiral Lagrangian for the mixed lattice theory with Wilson sea quarks and GW valence quarks was constructed. We first find all the relevant operators of  $\mcal O(a^{2})$ 
in  the  
Symanzik effective action \cite{Symanzik:1983dc,Symanzik:1983gh} for the mixed lattice theory. 
A spurion analysis similar to 
that in ordinary 
\chpt\ \cite{Weinberg:1979bn,Gasser:1984gg,Gasser:1983yg} is applied to the 
Symanzik effective action~\cite{Sharpe:1998xm, Rupak:2002sm}.
The result is a chiral effective theory 
that
exhibits explicit quark mass and lattice spacing dependence
of the underlying lattice theory. A recent review of this method can be found in Ref.\ \cite{Bar:2004xp}.

The leading order chiral Lagrangian presented here includes the lattice spacing 
effects proportional to $a^{2}$. Compared to the leading order Lagrangian of 
staggered \chpt\ \cite{Lee:1999zx,Bernard:2001yj,Aubin:2003mg,Aubin:2003uc}, which is the 
appropriate low-energy effective theory for ``unmixed'' lattice QCD with staggered sea and valence quarks, it contains only one additional operator together with an undetermined low-energy constant. 
This new operator contributes to the masses of ``mixed'' mesons (one valence and one sea quark) at
tree level in the chiral expansion.  The decay constant of valence-valence mesons then receives contributions
from the new operator through the mixed meson masses at one loop.  However, the masses of
valence-valence mesons themselves get no one-loop contribution from the new operator.
Besides an analytic contribution of $\mcal O(m_{\rm quark}a^{2})$, the one-loop pseudoscalar masses depend only on the leading order 
low-energy constants of staggered \chpt\ and the next-to-leading order (NLO)  
low-energy constants of continuum \chpt, the Gasser-Leutwyler coefficients.
To the extent that these low energy constants are known from previously performed 
staggered simulations \cite{Aubin:2004fs}, the quark mass and lattice spacing 
dependence of the valence-valence pseudoscalar mesons in the mixed 
theory are highly restricted and depend on only one free parameter.

The paper is organized as follows: In section~\ref{EffectiveLagrangian} we 
discuss  the Symanzik effective action for the mixed lattice 
theory and list 
all those operators that are relevant in the subsequent analysis. We then perform 
the necessary spurion analysis and derive the $\mcal O(a^{2})$ terms in the chiral effective 
Lagrangian.  The tree-level meson masses are written down in section~\ref{TreeLevel}.
In section~\ref{NLOMass} we calculate the valence-valence meson masses at one loop;
while the corresponding calculation of the pseudoscalar decay constants  is given
in section~\ref{NLODecayConstant}.
We conclude with a discussion of the results in section~\ref{summary}.
Appendix \ref{appendix1} is devoted to some technical details of our calculation.

%
\section{The chiral effective Lagrangian }
\label{EffectiveLagrangian}
%

%
\subsection{Lattice theory} 
%
Mixed fermion theories are a generalization of partially quenched theories. Theoretically they can be formulated by an action with anti-commuting sea and valence quarks and commuting valence ghosts \cite{Morel:1987xk}, where the quark masses and the Dirac operators are chosen differently in the sea and in the valence sector.\footnote{Instead of this ``ghost'' formulation one could also employ the ``replica method'' with valence quarks only \cite{Damgaard:2000gh}.} In the following we 
consider a mixed theory with $N_{f}$ staggered sea and $N_{V}$ valence fermions. 

The sea sector is described by the standard staggered fermion action, 
\bea\label{StaggeredAction}
S_{\rm Sea} & =& a^{4}\sum_{x,\mu}\overline{\chi}_{S}(x)[ \eta_{{\mu}}\nabla_{\mu}+ m_{\rm Sea}] \chi_{S}(x), 
\eea
where $\nabla_{\mu}$ denotes the gauge covariant central difference operator and $m_{\rm Sea}$ is the $N_{f}\times N_{f}$ mass matrix in the sea sector. For brevity we have suppressed the flavor and color indices. 
This action is invariant under lattice rotations, axis reversal, and translations by one lattice 
spacing (the so-called shift symmetry). In addition, the single-flavor staggered
theory possesses vector and (in the massless case)
axial-vector $U(1)$ symmetries. 
The explicit expressions for the field transformations that correspond to 
these symmetries can be found in Ref.\ \cite{Golterman:1984cy}.
For $N_f$ flavors, these symmetries extend to a $U(N_f)_\ell\times U(N_f)_r$ 
symmetry~\cite{Aubin:2003mg} that  
corresponds to flavor transformations on the
odd and even sites separately.\footnote{The subscripts $\ell$ and $r$ are used instead of $L$ and $R$ because the chiral rotations in $U(N_f)_\ell\times U(N_f)_r$ act on the spin {\em and} taste degrees of freedom (see Ref.\  \cite{Aubin:2003mg}).}

The action for the valence and ghost quarks is given by
\bea\label{SGW}
S_{\rm Val}  = a^4\sum\limits_x {\psibar _{V}(x) \left\{ {D_{GW}  
+ m_{\text{Val}}\,\left( {1 - \tfrac{1}
{2}aD_{GW} } \right) } \right\}\psi _{V} (x)}\,.
\eea
The valence and ghost quark masses are contained in the $2N_V\times 2N_V$ mass
matrix $m_\text{Val}$. 
The Dirac operator $D_{GW}$ is assumed to be a local operator  that satisfies the 
Ginsparg--Wilson relation~\cite{Ginsparg:1982bj},  
realized by either overlap \cite{Narayanan:1992wx,Narayanan:1993sk,Narayanan:1994gw} or domain-wall fermions \cite{Kaplan:1992bt}. Crucial is that the valence action has an exact chiral symmetry if the valence mass is set to zero \cite{Luscher:1998pq}. In addition it is invariant under the lattice symmetries (translations, rotations and reflections).
%
\subsection{Symanzik action}
\label{Symanzik}
%
At momenta $p$ much below the lattice cut-off momentum $1/a$, 
the long distance physics of the lattice theory can be described by the 
continuum Symanzik effective theory~\cite{Symanzik:1983dc,Symanzik:1983gh}. 
 The effects due to a non-zero lattice spacing appear in the form of higher dimensional operators in the effective action, multiplied by appropriate powers of $a$. These operators are constrained by the symmetries of the underlying lattice theory. The Symanzik effective action for the mixed lattice theory has the generic 
form 
\bea\label{StrucSymanzik}
S_{\rm Sym} & =&  S_{4} + a^{2} S_{6} +\cdots\ .
\eea
The first term is the known continuum partially quenched QCD action \cite{Morel:1987xk,Bernard:1993sv} containing  $4N_{f}$ sea quark fields $\overline{\psi}_{\rm S},{\psi}_{\rm S},$ and $N_{V}$ valence quark and ghost fields $\overline{\psi}_{\rm V}, {\psi}_{\rm V}$.\footnote{We collect a valence quark field $\psi_{V}^{\rm q}$ and the associated ghost field $\tilde{\psi}_{V}^{\rm gh}$ in one valence field $\psi_{V}=(\psi_{V}^{\rm q}, \tilde{\psi}_{V}^{\rm gh})$, and similarly for the anti-valence fields.} Each staggered flavor field comes in four different tastes, hence the four-fold degeneracy in the sea quark sector. The mass matrix consists of the renormalized quark masses proportional to the bare lattice quark masses. The symmetries of the lattice action forbid any dimension three operator that would lead to an additive mass renormalization. 

There are no terms linear in $a$ in eq.\ \pref{StrucSymanzik}, because no dimension five operator is compatible with the symmetries of the underlying lattice theory. 
Dimension five quark bilinears with two 
valence fields are ruled out by the chiral symmetry in the valence sector \cite{Niedermayer:1998bi}, and staggered quark bilinears are forbidden by the axial U(1) and the shift symmetry \cite{Sharpe:1993ng,Luo:1996vt}. Mixed bilinears with one sea and one valence field are not compatible with the separate flavor symmetries in the sea and valence sector.   

In order to find the terms in $S_{6}$ it will be convenient to distinguish three types of operators: Operators that involve only sea quark fields, operators that contain only valence quark fields and those that contain both.\footnote{We do not need to discuss purely gluonic operators, which also appear at 
$\mathcal O(a^{2})$, since they transform trivially under chiral transformations and therefore will not affect the form of the chiral effective Lagrangian.} The terms of the first two types have been constructed  previously and can be found in the literature. The operators involving only sea quark fields are listed in Refs.\ \cite{Luo:1997tt,Lee:1999zx} for the $N_{f}=1$ case, and the results were  generalized to the arbitrary $N_{f}$ case in 
Refs.\ \cite{Aubin:2003mg,Sharpe:2004is}. Similarly, the operators containing only valence fields are listed in Ref.\ \cite{Bar:2003mh}, where the Symanzik action for the mixed lattice theory with Wilson sea and GW valence quarks was constructed. 

What remains to be done here is the construction of the mixed operators containing sea and valence fields. Bilinears of dimension six with one sea and one valence field are ruled out by the separate flavor symmetries in the sea and valence sector, analogously to the dimension five bilinears.   
The mixed operators are therefore four-fermion operators that are products of two bilinears, one from the sea and one from the valence sector.

We construct these four-fermion operators by closely following the procedure described in Ref.\ \cite{Lee:1999zx}. We first construct all relevant lattice operators that are compatible with the symmetries of the lattice theory and correspond to dimension 6 four-fermion operators in the continuum limit. Sending then $a$ to zero gives the desired terms in the Symanzik action. This procedure was used in Ref.\ \cite{Lee:1999zx} to construct all four-fermion operators made of two staggered quark bilinears. The method is easily adapted to the mixed operators. We present the details of the construction in  Appendix \ref{appendix1} and quote here the final result. 

The allowed mixed four-fermion operators are of the form
\bea\label{allowedMixed4FOp}
O^{(6)}_{\rm Mix} & = & \overline{\psi}_{\rm S} (\gamma_{Spin}\otimes t^{a}_{Color}){\psi}_{\rm S}\, \overline{\psi}_{\rm V} (\gamma_{Spin}\otimes t^{a}_{Color}){\psi}_{\rm V}\,.
\eea
The matrix $\gamma_{Spin}$ represents one of the sixteen Clifford algebra elements and $t^{a}_{Color}$ denotes a color gauge group generator.\footnote{The identity in color space, for  which we use the notation $t_{Color}^{0}$, is also allowed here.} The Dirac and color indices are contracted in such a way that the four-fermion operator is a singlet under
$O(4)$ rotation symmetry and color. Furthermore, in writing eq.~\pref{allowedMixed4FOp} it is understood that the 
bilinear $\overline{\psi}_{\rm V} (\gamma_{Spin}\otimes t^{a}_{Color}){\psi}_{\rm V}$ is an $SU(N_{V}|N_{V})$  flavor singlet and that the 
bilinear $\overline{\psi}_{\rm S} (\gamma_{Spin}\otimes t^{a}_{Color}){\psi}_{\rm S}$ is an $SU(4N_{f})$ taste singlet. 
It is worth emphasizing that the mixed four-fermion operators do not break the taste symmetry in the sea quark sector. 

The separate axial symmetries in the sea and valence sector imply that the bilinears in eq.~\pref{allowedMixed4FOp} transform either as a vector or an axial  vector. Writing $\gamma_{Spin}\otimes t^{a}_{Color}$ more compactly as $\gamma_{S}t^{a}$ we therefore find only four mixed four-fermion operators that are allowed by the symmetries:
\begin{xalignat}{4}
  &{}&{}&O_{{\rm Mix},1}^{(6)}  = (\overline \psi_{S} \gamma _\mu  \psi_{S})(\overline \psi_{V} \gamma _\mu  \psi_{V}),&
     &O_{{\rm Mix},3}^{(6)}  = (\overline \psi_{S}  \gamma _\mu  t^a \psi_{S})(\overline \psi_{V}  \gamma _\mu t^a \psi_{V}),&{}&{}\label{MixedQuartic}\\
  &{}&{}&O_{{\rm Mix},2}^{(6)}  = (\overline \psi_{S} \gamma _\mu  \gamma _5 \psi_{S})(\overline \psi_{V} \gamma _\mu  \gamma _5 \psi_{V}),&
     &O_{{\rm Mix},4}^{(6)}  = (\overline \psi_{S}  \gamma _\mu  \gamma _5 t^a \psi_{S})(\overline \psi_{V}  \gamma _\mu  \gamma _5 t^a \psi_{V}),&{}&{}\notag
\end{xalignat}
where we have explicitly separated out bilinears containing the color identity matrix, and now restrict $t^a$
to be traceless,  summing over $a$.

\subsection{Spurion analysis}\label{spurionanalysis}

The leading term in the Symanzik action, $S_{4}$, is just the continuum action of partially quenched QCD. 
In the massless limit it is invariant under the flavor symmetry group
\bea\label{SymGroup}
G_{\rm PQ\,QCD} & =& SU(4N_{f} + N_{V}|N_{V})_{L} \otimes SU(4N_{f} + N_{V}|N_{V})_{R},
\eea
which is expected to be spontaneously broken to its vector part $SU(4N_{f} + N_{V}|N_{V})_{V}$.  
The low-energy physics is therefore dominated by Nambu-Goldstone bosons. These pseudoscalar bosons acquire small masses due to  non-vanishing quark masses and a non-zero lattice spacing. The latter contribution has its origin in chiral symmetry breaking terms in $S_{6}$. 

To construct the chiral Lagrangian that describes these pseudoscalar bosons we follow the standard procedure of a spurion analysis. The coefficient $c_{i}$ of each term $O_{i}$ in the Symanzik effective action is promoted to a spurion field that transforms under flavor transformations in eq.~\pref{SymGroup} in such a way that the product $c_{i}O_{i}$ is invariant. The chiral effective Lagrangian is constructed from the pseudoscalar fields and the spurion fields with the requirement that it is invariant under flavor rotations. Once the chiral Lagrangian is constructed the spurion field is set to its original constant value. This 
guarantees
that the chiral Lagrangian explicitly breaks the chiral symmetries in the same manner as the underlying Symanzik effective action, and reproduces the same Ward identities. 

In order to perform the spurion analysis for the mixed theory it is convenient to introduce the following notation. We collect the quark and anti quark fields in single fields,
\bea
\Psi = (\psi_S,\psi_V),\qquad \overline{\Psi} =  (\overline{\psi}_S,\overline{\psi}_V),
\eea
where $\psi_V$ contains both the anticommuting valence quarks and the commuting ghost fields. 
The mass matrix is given by
$m=\operatorname{diag}(m_S,m_V)$, with $m_S$ being the $4N_f\times 4N_f$
mass matrix for  the sea quarks and $m_V$ is the $2N_V\times 2N_V$ mass matrix for the valence quarks and valence  ghosts. We also introduce the projection operators
\be
P_S=\operatorname{diag}(I_{S},0)\,,\qquad P_V=\operatorname{diag}(0,I_{V})\,,
\ee
where $I_{S}$ denotes the $4N_{f}\times 4N_{f}$ identity matrix in the
sea sector, and $I_{V}$ the $2N_{V}\times 2N_{V}$ identity matrix in
the space of valence fields.

For our purposes it will not be necessary to construct all spurion fields that make the Symanzik action in eq.\ \pref{StrucSymanzik} invariant. Most of the analysis has already been done and we can rely on previously published results. 
The results of the spurion analysis  to $\mcal O(a^2)$
for the case with staggered sea and valence quarks were written down in Ref.~\cite{Aubin:2003mg}; the
analysis can be found in detail
in Ref.\ \cite{Sharpe:2004is}, which also works to higher order.
  Since now only the 
 sea sector contains staggered quarks all we need to do is to include 
 the projector $P_{S}$ appropriately in all spurion fields 
associated with sea quarks in this reference. These spurion fields render invariant all terms in the Symanzik 
effective action that are built only of sea quarks.
To illustrate this point consider the mass term for the sea quarks, $\overline{\psi}_{S}m_{S}\psi_{S}$. In 
Ref.\ \cite{Sharpe:2004is} this term is made invariant by promoting the mass to a spurion field that transforms as $m_{S} \rightarrow Lm_{S}R^{\dagger}$ under left- and right transformations $L$ and $R$. In order to make use of this spurion field in the mixed theory we write $\overline{\psi}_{S}m_{S}\psi_{S} = \overline{\Psi} P_{S}mP_{S}\Psi$ and assume the same transformation property as before, \ie $m \rightarrow L m R^{\dagger}$.  The constant value to which the spurion is assigned in the end, however,  is now $P_{S}mP_{S}$. One can proceed analogously with all the other spurion fields in Ref.\ \cite{Sharpe:2004is}.
From the results in Ref.\ \cite{Bar:2003mh} for the mixed theory with Wilson sea quarks and GW valence quarks we can directly  determine 
the spurion fields that are necessary to make the valence field operators invariant.  What we need in addition are the new spurion fields that make the mixed four-fermion operators in eq.\ \pref{MixedQuartic} invariant. 

We want to emphasize that the mixed four-fermion operators do break the symmetry group $G_{\rm PQ\,QCD}$, even though each bilinear in them couples fields of the same chirality only. The reason is that any four-fermion operator that is invariant under all transformations in $G_{\rm PQ\,QCD}$ must be of the form
\bea\label{StructureInvFF}
(\overline{\Psi}\Gamma\Psi)^{2} & =& (\psi_{S}\Gamma\psi_{S})^{2} + (\psi_{V}\Gamma\psi_{V})^{2} +2 (\psi_{S}\Gamma\psi_{S})(\psi_{V}\Gamma\psi_{V}),
\eea
where $\Gamma$ represents one of the four combinations $\gamma_{S}t^{a}$ in eq.\ \pref{MixedQuartic}. All three types of four-fermion operators on the right hand side of eq.~\pref{StructureInvFF} appear in the Symanzik effective action. However,  because the lattice theory does not posses any  symmetries relating the staggered sea and the GW valence fermions, they do not enter with the same coefficient in front in order to sum up to the square on the left hand side. Consequently, although an arbitrary linear combination of the three operator types is invariant under the subgroup $SU(4N_{f})_{L}\otimes SU(4N_{f})_{R}\otimes SU(N_{V}|N_{V})_{L} \otimes SU(N_{V}|N_{V})_{R}$ of $G_{\rm PQ\,QCD}$, it is not invariant under  the larger symmetry transformations of $G_{\rm PQ\,QCD}$ 
that mix the sea and valence sector.

Following the notation in Ref.\ \cite{Bar:2003mh}, the mixed four-fermion operators can be made invariant under arbitrary  chiral flavor transformations $L \in SU(4N_{f} + N_{V}|N_{V})_L$ and $R \in SU(4N_{f} + N_{V}|N_{V})_R$ by introducing the spurion fields 
\begin{align}
&\qquad\begin{aligned}\label{mixedspurions} 
& D \equiv D_1  \otimes D_2  \to LD_1 L^\dag   \otimes LD_2
   L^\dag, & \quad E \equiv E_1  \otimes E_2  \to RE_1 R^\dag   \otimes
   RE_2 R^\dag ,  \\  
   & F \equiv F_1  \otimes F_2  \to LF_1 L^\dag   \otimes RF_2 R^\dag,
   & \quad G \equiv G_1  \otimes G_2  \to RG_1 R^\dag   \otimes LG_2
   L^\dag , 
\end{aligned}\\  
&\qquad D_0 =E_0 =F_0 =G_0  =a^2P_S\otimes P_V.\nn
\end{align}
The constant values to which the spurion fields are assigned to in the end carry the subscript ``0''. The spurion fields transform as 4-tensors under chiral flavor transformations and therefore carry four indices, which need to be properly contracted with the indices of the fermion fields in order to form invariants.\footnote{Spurion fields with the same transformation properties appear also in weak matrix element studies. See Ref.\ \cite{Bernard:1989nb} and references therein.} 
For example, decomposing $O_{{\rm Mix},1}^{(6)} = (\overline \psi_{S} \gamma _\mu  \psi_{S})(\overline \psi_{V} \gamma _\mu  \psi_{V})\equiv(\overline \psi\gamma _\mu  \psi)_{S}(\overline \psi\gamma _\mu  \psi)_{V}$ in chiral components we obtain
\bea\label{ChiralDecompO6}
(\overline \psi\gamma _\mu  \psi)_{S}(\overline \psi\gamma _\mu  \psi)_{V}& = & \phantom{+}(\overline \psi_{L}\gamma _\mu  \psi_{L})_{S}(\overline \psi_{L}\gamma _\mu  \psi_{L})_{V}  +(\overline \psi_{L}\gamma _\mu  \psi_{L})_{S}(\overline \psi_{R}\gamma _\mu  \psi_{R})_{V}\\
& & + (\overline \psi_{R}\gamma _\mu  \psi_{R})_{S}(\overline \psi_{L}\gamma _\mu  \psi_{L})_{V} + (\overline \psi_{R}\gamma _\mu  \psi_{R})_{S}(\overline \psi_{R}\gamma _\mu  \psi_{R})_{V}.\nn
\eea
Both bilinears in the first term on the right hand side couple left-handed fields only. It is made invariant with the spurion field $D$, where the indices are contracted according to 
\bea
D(\psi_{L}\gamma _\mu  \psi_{L})_{S}(\overline \psi_{L}\gamma _\mu  \psi_{L})_{V}  & =& (\psi_{L}D_{1}\gamma _\mu  \psi_{L})_{S}(\overline \psi_{L}D_{2}\gamma _\mu  \psi_{L})_{V} .
\eea
The other three terms in eq.\ \pref{ChiralDecompO6} are analogously made invariant using the spurion fields  $E,F$ and $G$.

We remark that the spurion fields in eq.\ \pref{mixedspurions} transform in the same way as the ones for the valence-valence four-fermion operators; they differ only in their constant final values: $a^2P_S\otimes P_V$ is replaced by $a^2P_V\otimes P_V$ in the valence-valence case~\cite{Bar:2003mh}.

\subsection{Chiral Lagrangian}
Assuming that the symmetry in eq.\ \pref{SymGroup} is spontaneously broken down to its vector part, the particle spectrum contains light pseudoscalar bosons. These bosons are described by the field 
\begin{equation}\eqn{SigmaDef}
\Sigma=\exp(2i\Phi / f) \,,
\end{equation}
which is an element of $U(4N_{f} + N_{V}|N_{V})$.
$\Phi$ is a matrix that collects the pseudoscalar fields in the usual
way \cite{Bernard:1993sv}.
For example, for three sea quark 
flavors ($u$, $d$ and $s$)  and two valence flavors ($x$, $y$ 
for the valence quarks and $\tilde{x}$, $\tilde{y}$ for the valence ghosts) 
we arrange the fields as follows:
\begin{eqnarray}\eqn{PhiDef}
\Phi = \left( \begin{array}{cccccccc} 
 U & \pi^+ & K^+&  Q_{ux} &  Q_{uy} &  \cdots & \cdots \\
 \pi^- & D & K^0& Q_{dx} & Q_{dy} & \cdots & \cdots \\*
 K^- & \bar{K^0} & S &  Q_{sx} & Q_{sy} & \cdots & \cdots \\*
 Q_{ux}^\dagger & Q_{dx}^\dagger & Q_{sx}^\dagger &  X & P^+ & R_{\tilde xx}^\dagger	& R_{\tilde yx}^\dagger \\* 
 Q_{uy}^\dagger  & Q_{dy}^\dagger  & Q_{sy}^\dagger   &  P^- & Y &R_{\tilde xy}^\dagger	& R_{\tilde yy}^\dagger	 \\*
 \cdots & \cdots  & \cdots  & R_{\tilde xx} & R_{\tilde xy}  & \tilde{X} & \tilde{P}^+ \\*
 \cdots & \cdots  & \cdots  & R_{\tilde yx}  & R_{\tilde yy}  & \tilde{P}^- & \tilde{Y} \\*
  \end{array}
  \right).\end{eqnarray}
Here $P^+$, $X$ and $Y$ are the $x\bar y$, $x\bar x$, and $y \bar y$ valence bound
states, respectively; $\tilde P^+$, $\tilde X$ and $\tilde Y$ are the analogous combinations
of valence ghost quarks.
$R_{\tilde x x}$ is the (fermionic) bound state 
$\tilde x\bar x$, and similarly for $R_{\tilde xy}$, $R_{\tilde yx}$, and $R_{\tilde yy}$.
Likewise, $Q_{Fv}$ represents
the mixed bound state $F\bar v$, where $F$ is a sea quark, $F\in\{u,d,s\}$, and
$v$ is a valence quark, $v\in\{x,y\}$.\footnote{We have not bothered to 
name the mixed bound states of sea quarks 
with ghost valence quarks in \protect{\eq{PhiDef}} because such 
states will not enter into the calculations below.}
$Q_{Fv}$ is a $4\times 1$ matrix in taste; 
while $Q_{Fv}^\dagger\equiv Q_{vF}$
is a $1\times4$ matrix. 
The sea-quark bound state fields are $U$, $\pi^{+}$, $K^{+}$, {\it etc.} 
These are $4\times4$ 
matrices when we take into account the taste degree of freedom. 
We write
\bea\eqn{UDef}
U&=&\sum_{b=1}^{16}U_{b}\frac{T_{b}}{2}\ ,
\eea
(and similarly for $\pi^{+},K^{+},\ldots$) where 
\bea
T_b = \{ \xi_5, i\xi_{\mu}\xi_{5}, i\xi_{\mu}\xi_{\nu}, \xi_{\mu},\xi_I\}
\eea
are the sixteen taste matrices in the form of Euclidean gamma matrices 
($\xi_I$ denotes the $4\times4$ identity matrix).
Unlike Ref.~\cite{Aubin:2003mg}, we include a factor of 2 in \eq{SigmaDef}
and divide $T_b$ by 2 in \eq{UDef} in order to keep a consistent normalization
of all fields in \eq{PhiDef}.

Under  chiral symmetry transformations in eq.~\pref{SymGroup}, the field $\Sigma$
transforms as
\begin{equation}
\Sigma \rightarrow  L \Sigma R^{\dagger}\,,
\label{eq:sigtrans}
\end{equation}
where $L \in SU(4N_{f} + N_{V}|N_{V})_L$ and $R \in SU(4N_{f} + N_{V}|N_{V})_R$.

The chiral Lagrangian is expanded in powers of $p^{2}$, $m_{q}$ and $a^{2}$, where $m_{q}$ stands generically for either a sea or a valence quark mass. We adopt a power counting that assumes that the size of the chiral symmetry breaking due to the quark masses and the discretization
effects are of comparable size, \ie
\begin{equation}\label{powercounting}
m_{q}/\Lambda_{QCD} \approx a^2
\Lambda_{QCD}^2,
\end{equation}
where $\Lambda_{QCD}$ denotes the typical QCD scale, of the order of 300 MeV.
A different power counting is necessary if one of the two parameters $m_{q}/\Lambda_{QCD}$ and $a^2\Lambda_{QCD}^2$ is much larger than the other one. However, the approximate equality in  eq.~\pref{powercounting} is realized in  current lattice simulations using improved staggered fermions \cite{Aubin:2004fs}. 

Assuming eq.~\pref{powercounting}, the leading order chiral 
Lagrangian contains the terms of $\mcal O(p^2,m_q,a^{2})$
and is of the form
\begin{equation}\eqn{LOChiralLagrangian}
	{\cal L}_{\chi} = \frac{f^2}{8} \langle\partial_{\mu}\Sigma
	\partial_{\mu}\Sigma^{\dagger}\rangle - \frac{f^2B}{4}\langle 
	\Sigma M^{\dagger} + M \Sigma^{\dagger}\rangle + \frac{m_0^2}{6}\langle \Phi \rangle^2
        + a^2 {\cal V}.
\end{equation}
Here $\langle \ldots \rangle$ denotes a supertrace in flavor space and the 
parameters $f$ and $B$ are undetermined low-energy constants.\footnote{We adopt 
a normalization that corresponds to a tree-level pion decay constant $f\approx 131$ MeV.}  
For our concrete example of three sea quark flavors and two valence flavors the diagonal 
mass matrix $M$ is given by 
$M=$diag$(m_u \xi_{I},m_d \xi_{I},m_s \xi_{I},m_{x},m_{y},m_{x},m_{y})$.

As in Ref.~\cite{Bernard:2001yj,Aubin:2003mg,Aubin:2003uc},
for convenience we leave explicit the $m_0^2$ term,
which is allowed because of the anomaly, and we take
 $m_0^2\to\infty$ at the end.
Note that
\begin{equation}\eqn{PhiI}
\langle \Phi \rangle = 2U_I + 2D_I + 2S_I + X +Y -\tilde X -\tilde Y \ ,
\end{equation}
where $U_I$ is the taste singlet component of $U$ (\eq{UDef}), and similarly for
$D_I$ and $S_I$.

The potential ${\cal V}$ in the leading order Lagrangian comprises all terms proportional to $a^{2}$. For our mixed theory it can be written as a sum of three terms,
\begin{equation}
{\cal V} = {\cal U}_{\rm S} +{\cal U}^{\prime}_{\rm S}  + {\cal U}_{\rm V}.
\end{equation}
The first two terms are just the known taste breaking potentials for the sea sector \cite{Lee:1999zx,Aubin:2003mg}:
\begin{align}\eqn{US}
-{\cal U}_{S}  = & 
\phantom{+}C_1	\langle \hat{\xi}_5P_{S}\Sigma\,\hat{\xi}_5P_{S}\Sigma^{\dagger}\rangle 
+C_3\frac{1}{2} \sum_{\nu} \Big[\langle \hat{\xi}_{\nu}P_{S}\Sigma \,\hat{\xi}_{\nu}P_{S}\Sigma\rangle   \,+\, \mbox{h.c.}\Big] 
\\ & 
+C_4\frac{1}{2} \sum_{\nu}\Big[\langle \hat{\xi}_{\nu 5}P_{S}\Sigma \,\hat{\xi}_{5\nu}P_{S}\Sigma\rangle \,+\, \mbox{h.c.}
\Big]\nn
+C_6 \sum_{\mu<\nu} \langle \hat{\xi}_{\mu\nu}P_{S}\Sigma\, \hat{\xi}_{\nu\mu}P_{S}\Sigma^{\dagger}\rangle\nn
\end{align}
and
\begin{align}\eqn{UprimeS}
-{\cal U}^{\prime}_{S} = & 
\phantom{+}C_{2V}\frac{1}{4} \sum_{\nu}\Big[\langle\hat{\xi}_{\nu}P_{S}\Sigma\rangle
\langle\hat{\xi}_{\nu}P_{S}\Sigma\rangle \,+\, \mbox{h.c.}\Big]
+C_{2A}\frac{1}{4} \sum_{\nu}\Big[\langle\hat{\xi}_{\nu 5}P_{S}\Sigma\rangle\langle\hat{\xi}_{5\nu}P_{S}\Sigma\rangle \,+\, \mbox{h.c.}
\Big] \\ &
	+C_{5V}\frac{1}{2} \sum_{\nu}\langle\hat{\xi}_{\nu}P_{S}\Sigma\rangle
	\langle\hat{\xi}_{\nu}P_{S}\Sigma^{\dagger}\rangle \nn 
	+C_{5A}\frac{1}{2} \sum_{\nu}\langle\hat{\xi}_{\nu5}P_{S}\Sigma\rangle
	\langle\hat{\xi}_{5\nu}P_{S}\Sigma^{\dagger}\rangle.
\end{align}
Here we introduced the short hand notation $\hat{\xi}_5  = (\xi_{5,{\rm taste}}\otimes 1_{\rm flavor} ) \oplus 1_{V}$, etc., for the multi-flavor generalizations of the taste matrices $T_{b}$. For instance, in our concrete example we have $\hat{\xi}_5 = \mbox{diag}(\xi_{5}, \xi_{5},\xi_{5},1,1,1,1)$. The coefficients $C_{i}$ are low-energy constants.

Note that all terms in the two potentials ${\cal U}_{\rm S}$ and 
${\cal U}^{\prime}_{\rm S}$ 
only involve the fields in the upper left sea-sea block of $\Sigma$. This is easily seen by first noting that the matrices $\hat{T}_{b}$ commute with the projector $P_{S}$. Consequently, the structure $\langle {T}_{b}P_{S}\Sigma {T}_{b}P_{S}\Sigma^{\dagger}\rangle$, for instance,  can also be written as $\langle {T}_{b}P_{S}\Sigma P_{S} {T}_{b}P_{S}\Sigma^{\dagger}P_{S}\rangle$. 
This is not surprising. The taste matrices $\hat{T}_{b}$ played the role of spurion fields in the derivation of the potential ${\cal U}_{\rm S} +{\cal U}^{\prime}_{\rm S}$ in Ref.\ \cite{Aubin:2003mg,Sharpe:2004is}. As explained in section \ref{spurionanalysis}, the spurion fields need to be sandwiched by the projector $P_{S}$ in the mixed theory, \ie $\hat{T}_{b} \rightarrow P_{S}\hat{T}_{b}P_{S}$.  In fact, that was the way we obtained the potential ${\cal U}_{\rm S} +{\cal U}^{\prime}_{\rm S}$ without repeating the details of the spurion analysis in Ref.\ \cite{Aubin:2003mg,Sharpe:2004is}.  

The remaining potential ${\cal U}_{\rm V}$ comprises all terms that stem from the operators in the Symanzik effective action that involve valence fields. Some of these arise from the mixed 4-fermion operators in
eq.~(\ref{MixedQuartic}). The corresponding chiral operators
 are constructed by forming invariants involving one of the spurion fields in 
eq.~\pref{mixedspurions} together with arbitrarily many $\Sigma$ and $\Sigma^{\dagger}$. Insertions of the mass matrix and derivatives are excluded since they lead to terms that are necessarily of higher order in the chiral expansion, at least $\mcal O(p^{2}a^{2}, m_{q}a^{2})$. 

We can only form non-trivial invariants with the fields $F$ and $G$, which collapse to the same term  once the spurion fields are set to their final value:
\beq
 &&\left\langle {F_1 \Sigma F_2 \Sigma ^\dag  } \right\rangle  \to a^2
 \left\langle {\tau _3 \Sigma \tau _3 \Sigma ^\dag  } \right\rangle ,
 \eqn{Finvariant}\\   
 &&\left\langle {G_1 \Sigma^{\dagger} G_2 \Sigma  } \right\rangle  \to a^2
 \left\langle {\tau _3 \Sigma^{\dag} \tau _3 \Sigma  } \right\rangle.\nn
 \eqn{Ginvariant}
\eea
Here we used  $P_S=\tfrac{1}{2}(I+\tau_3)$ and
$P_V=\tfrac{1}{2}(I-\tau_3)$, with
$\tau_3=\operatorname{diag}(I_{S},-I_{V})$ and dropped an irrelevant factor 
of 1/4. When we write
$(I\pm\tau_3)$ for $F_{1,2}$ and $G_{1,2}$, the fields $\Sigma$ and $\Sigma^\dag$ cancel whenever they 
sandwich the identity matrix. The only non-trivial operator is the one involving two $\tau_{3}$ matrices.

The last terms in ${\cal U}_{\rm V}$ stem from the four-fermion operators involving only valence fields.\footnote{The allowed 
valence bilinears either have the same symmetries as the lowest order terms, in which case they
merely give $\mcal O(a^2)$ corrections to lowest order parameters, or they have different symmetries (violate $O(4)$ rotation invariance) and
contribute only at higher order in the chiral expansion \cite{Bar:2003mh}.}  
As we remarked at the end of the last section, the corresponding spurion fields transform exactly as the ones for the mixed four-fermion operators. The only difference is the final constant value; $a^{2}P_{S}\otimes P_{V}$ is replaced by $a^{2}P_{V}\otimes P_{V}$. This change only involves a sign flip in the first projection operator, and therefore leads to the same term $a^2 \left\langle {\tau _3 \Sigma \tau _3 \Sigma ^\dag  } \right\rangle$ for the chiral Lagrangian.\footnote{This also explains that this term is present
independently of a change of basis in the Symanzik effective action. Using eq.\ \pref{StructureInvFF} one could replace either the mixed or the pure valence four-fermion operators at the expense of  $(\overline{\Psi}\Gamma\Psi)^{2}$, which is invariant under transformations in $G_{\rm PQ\,QCD}$. Nevertheless, one can remove only one type of four-fermion operators, the other type still gives rise to the term $a^2 \left\langle {\tau _3 \Sigma \tau _3 \Sigma ^\dag  } \right\rangle$ in the chiral Lagrangian.}

We conclude that the leading order chiral Lagrangian for the mixed action theory with staggered sea and 
GW valence quarks contains only one more operator compared to the chiral Lagrangian of \schpt:
\begin{align}\eqn{LsquareAdd}
\mcal U_{\rm V} &= -C_{\rm Mix} \left\langle
  \tau_3\Sigma\tau_3\Sigma^\dag\right\rangle.
  \end{align}
The potential ${\cal V}$ for the mixed action theory involves 
nine unknown low-energy constants compared to eight in \schpt. We want to emphasize that the eight constants in ${\cal U}_{\rm S} +{\cal U}^{\prime}_{\rm S}$ are the {\em same} constants as those in \schpt. Some combinations of them have 
already been determined by fits to staggered lattice data \cite{Aubin:2004fs}.

The potential ${\cal U}_{\rm S} +{\cal U}^{\prime}_{\rm S}$ breaks the $SU(4)$ taste symmetry in the sea sector but not entirely -  
 an accidental $SO(4)$ subgroup remains unbroken \cite{Lee:1999zx}. The part ${\cal U}_{\rm V}$, on the other hand,  preserves the full $SU(4)$ taste symmetry. This is expected because the four-fermion operators in eq.\ \pref{MixedQuartic}, which give rise to ${\cal U}_{\rm V}$, are trivial in taste space. 
Interaction vertices involving pseudoscalars 
with one or more valence-quark constituents stem from ${\cal U}_{\rm V}$ only. 
The $SU(4)$ taste symmetry implies that correlation functions that
include such external pseudoscalars respect the $SU(4)$ taste symmetry in
one-loop NLO diagrams. Analytic, taste symmetry violating contributions do appear at NLO: $\mcal O(a^4,a^2p^2,a^2m_q)$; while non-analytic symmetry violating contributions start beyond NLO.

\section{Pseudoscalar masses and decay constants}

In this section we compute the one-loop expressions for the (valence-valence) pseudoscalar masses and decay constants. For concreteness we now restrict ourselves to the case with three sea quark flavors and two valence quark flavors. 
This is the most relevant case phenomenologically.
Further, the resulting one-loop expressions can 
be readily used in the analysis of unquenched configurations generated with the Asqtad action by the MILC collaboration (see Refs.\ \cite{Aubin:2004wf,Aubin:2004fs} and references therein). 

In the following we also adjust for the so-called ``fourth root trick'', which is commonly employed in staggered simulations in order to reduce the taste degree of freedom from four to one. In the context of the chiral effective theory this adjustment requires proper insertions of factors of $1/4$ in the sea quark loop contributions in our expressions \cite{Bernard:2001yj}, depending on the quark flow \cite{Sharpe:1992ft} that corresponds to the meson loop diagram in the chiral effective theory.  

The fourth root trick raises legitimate locality questions and its validity is controversial. 
Recently various studies have addressed this issue using either numerical or analytical 
methods \cite{Durr:2003xs,Durr:2004as,Durr:2004rz,Durr:2004ta,Follana:2004sz,Maresca:2004me,Adams:2004mf,Shamir:2004zc}.
In the following we assume that the ``fourth root trick'' can be given a field theoretically sound underpinning, so that we can follow the procedure described in Ref.\ \cite{Bernard:2001yj}.  

\subsection{Leading-order masses and propagators}
\label{TreeLevel}

At tree level, the new operator of the mixed theory,
${\cal U}_V$, contributes only to the masses of 
valence-sea mesons, represented in 
\eq{PhiDef} by $Q_{Fv}$ ($F\in\{u,d,s\}$; $v\in\{x,y,\tilde x, \tilde y\}$). 
Sea-sea and valence-valence mesons get no
such contributions because a block-diagonal $\Phi$ commutes
with $\tau_3$, and $\Sigma\Sigma^\dagger=I$.  Similarly, the potentials
in the sea sector, ${\cal U}_S$ and ${\cal U}'_S$, \eqs{US}{UprimeS}, give
no contribution to the valence-valence mesons. Thus the
valence-valence mesons obey the continuum-like mass relations
exemplified by
\begin{equation}
\eqn{VVmass}
	m^2_{P} = B(m_x + m_y)\ .
\end{equation}
Of course  such relations follow immediately from the exact chiral symmetry (for massless
quarks) in the valence
sector.

In the sea-sea sector, ${\cal U}_S$ and ${\cal U}'_S$ contribute, and the tree-level results
are identical to those in Ref.\ \cite{Aubin:2003mg}.  For a  meson of taste $b$ made up
of sea quarks $F$ and $F'$ ($F\not=F'$), we have
\begin{equation}\label{eq:SSmass}
        m^2_{FF',b} = B (m_F + m_F') + a^2\Delta(\xi_b),
\end{equation}
with
\begin{eqnarray}\label{eq:deltas}
        \Delta (\xi_5) & \equiv & \Delta_P = 0\,,
                \\*
        \Delta (\xi_{\mu5}) & \equiv & \Delta_A = \frac{16}{f^2}\left(
        C_1 + 3C_3 + C_4 + 3C_6 \right)\,, \nonumber \\*
        \Delta (\xi_{\mu\nu})  & \equiv &\Delta_T =
                \frac{16}{f^2}\left(2C_3 + 2C_4 + 4C_6\right)\,, \nonumber \\*
        \Delta (\xi_{\mu}) & \equiv & \Delta_V = \frac{16}{f^2}\left(
        C_1 + C_3 + 3C_4 + 3C_6 \right)\,, \nonumber \\*
        \Delta (\xi_I)  & \equiv & \Delta_I =
         \frac{16}{f^2}\left(
        4C_3 + 4C_4 \right).\nn
\end{eqnarray}

As mentioned above, ${\cal U}_V$ contributes to the masses of the valence-sea mesons.
For the $F\bar x$ meson, with field $Q_{Fx}$, the mass is
\begin{equation}
\eqn{VSmass}
	m^2_{Fx} = B(m_F + m_x) + a^2\Delta_{\rm Mix}\ ,
\end{equation}
where
\begin{equation}
\eqn{DeltaMix}
	\Delta_{\rm Mix} \equiv \frac{16 C_{\rm Mix}}{f^2}\ .
\end{equation}
The violation of the Goldstone theorem in \eq{VSmass} clearly arises because
there is no lattice axial symmetry 
that mixes valence and sea quarks.

In a simulation using staggered sea quark configurations, $\Delta_{\rm Mix}$ 
could be directly determined from the
propagator of a mixed meson with one GW valence quark and one 
staggered valence quark.\footnote{At least within the context of 
staggered chiral perturbation theory,
taking the fourth root of the staggered sea-quark
determinant will not change the standard equivalence between masses of particles
on internal and external lines.  The fourth-root procedure only changes the weighting,
not the mass, of an internal meson made from one valence and one sea quark.}
Such a direct determination of $\Delta_{\rm Mix}$ would be useful because it 
would reduce the number of free parameters in chiral-log fits.  For example, 
$\Delta_{\rm Mix}$ enters (through $m^2_{fx}$) into the NLO expression for the
decay constant of a meson with two GW valence quarks
(see section \ref{NLODecayConstant}).  

Since we have no {\it a priori}\/ reason to expect a particular sign for $C_{\rm Mix}$
(or equivalently $\Delta_{\rm Mix}$), \eq{VSmass} shows
there is a possibility of a lattice ``Aoki phase'' \cite{Aoki:1983qi}  if $\Delta_{\rm Mix}< 0$.
This would be similar to the type of lattice phases for staggered quarks discussed in 
Refs.~\cite{Lee:1999zx,Aubin:2003mg,Aubin:2004dm}.
The direct measurement of $\Delta_{\rm Mix}$ discussed in the previous paragraph
would determine if this possibility is realized in practice.

\begin{figure}[tbh]
\resizebox{4.0in}{!}{\includegraphics{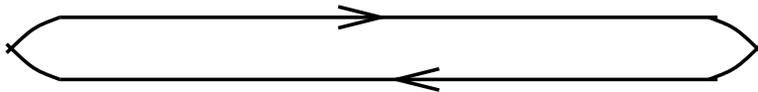}}
\caption{
Quark flow diagram for a connected meson propagator.
\label{fig:ConnProp}}
\end{figure}

\begin{figure}[tbh]
\resizebox{5.0in}{!}{\includegraphics{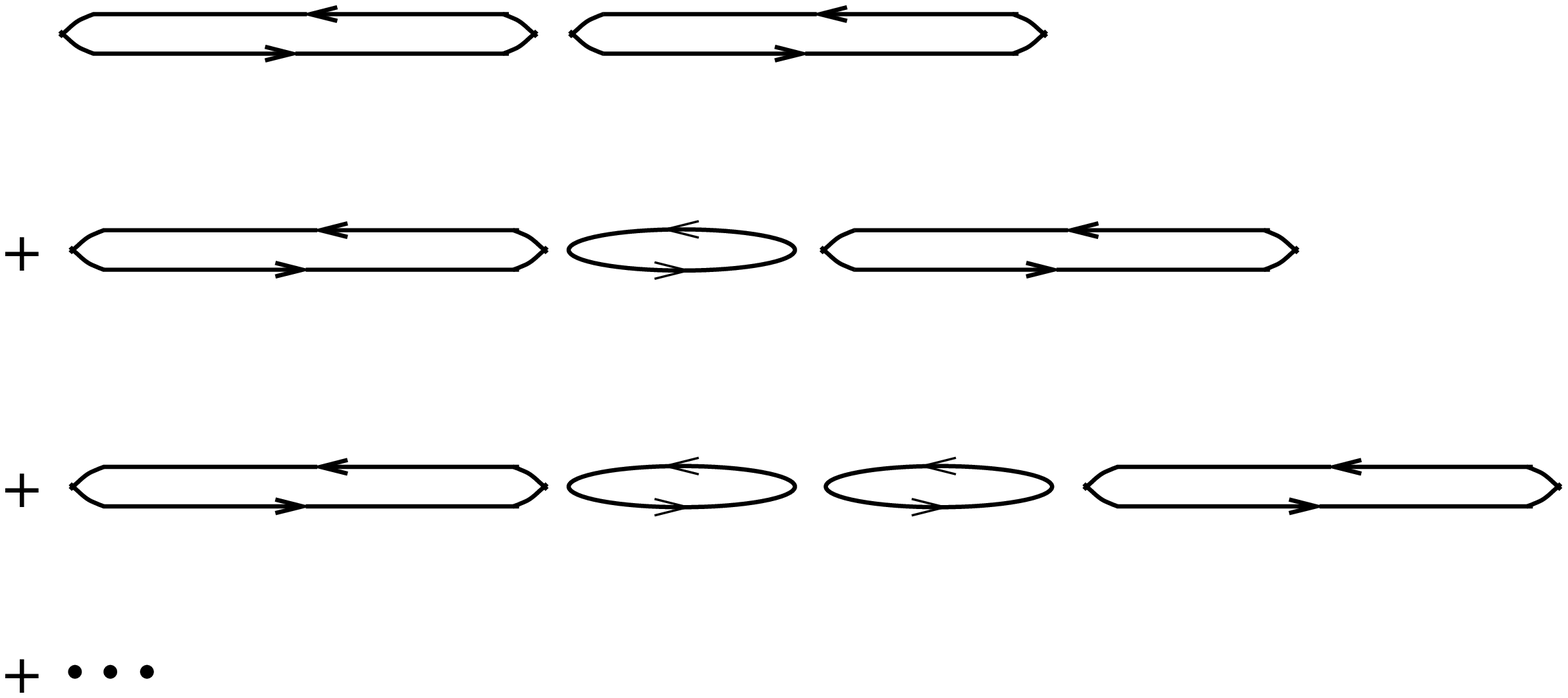}}
\caption{
Quark flow diagrams for the disconnected meson propagator.
\label{fig:DiscProp}}
\end{figure}

The flavor-charged (non-diagonal) fields in \eq{PhiDef} have only connected propagators
in the quark-flow sense, \figref{ConnProp}; while flavor neutral (diagonal, \eg $U$ or $X$) 
mesons also have disconnected contributions, \figref{DiscProp}.
The only complication in the connected case is getting
the sign of the propagator right for mesons with one or more ghost valence quarks.
Using $[AB](p^2)$ to denote the Euclidean space propagator of field $A$ and $B$ 
with momentum $p$, examples of connected
propagators are
\begin{eqnarray}\eqn{ConnProp}
\[P^+ P^-\] (p^2)& = & \frac{1}{p^2 + m^2_{P}}, \\
\[\pi^{+}_a \pi^{-}_b\] (p^2)& = & \frac{\delta_{ab}}{p^2 + m^2_{\pi_a}} \nonumber, \\
\[X X\]_{\rm conn} (p^2)& = & \frac{1}{p^2 + m^2_{X} }\nonumber, \\
\[\tilde X \tilde X\]_{\rm conn} (p^2)& = & \frac{-1}{p^2 + m^2_{\tilde X} }\nonumber, \\
\[Q_{u_ix}Q^\dagger_{u_jx}\] (p^2)& = & \frac{\delta_{ij}}{p^2 + m^2_{ux} }\nonumber, \\
\[R_{\tilde xy}R^\dagger_{\tilde xy}\] (p^2)& = & 
-\[R^\dagger_{\tilde xy}R_{\tilde xy}\] (p^2)= \frac{1}{p^2 + m^2_{\tilde xy} }\ .\nn
\end{eqnarray}
Here $a,b=\{1,2,\dots,16\}$ are meson taste indices as in \eq{UDef}; while
$i,j=\{1,2,3,4\}$ are quark taste indices.

Disconnected propagators for flavor neutral mesons can be generated by
the anomaly ($m_0^2$) term in \eq{LOChiralLagrangian}, which
gives a ``hairpin'' interaction (two-meson vertex with disconnected quark flow). 
Summing such graphs as in Refs.~\cite{Sharpe:2000bc,Aubin:2003mg} gives, for the disconnected
$X$-$Y$ propagator
\begin{equation}\label{eq:DiscPropPrime}
        \[XY\]_{\rm disc}(p^2)=
-\frac{ m^2_0}{3}\  \frac{(p^2 + m_{U_I}^2)(p^2 + m_{D_I}^2)(p^2 + m_{S_I}^2)}
                { (p^2 + m_{X}^2)(p^2 + m_{Y}^2) (p^2 + m_{\pi^0_I}^2)(p^2 + m_{\eta_I}^2) (p^2 + m_{\eta'_I}^2)}
\ ,
\end{equation}
where, for concreteness, we have again assumed three sea-quark flavors.  The $\pi^0_I$,
$\eta_I$ and $\eta'_I$ are the mass eigenstates in the flavor-neutral,
taste-singlet channel, found by diagonalizing the mass matrix including the $m_0^2$
term.  The subscript ``disc'' is included for clarity, but of course the $XY$ propagator
has no connected contribution. 
In the sea-sea sector, only the taste-singlet,
flavor neutral mesons feel the anomaly hairpin vertex.
In addition, sea-sea flavor neutral mesons of vector or axial taste get hairpin
contributions from ${\cal U}'_S$ \cite{Aubin:2003mg}. Since the corresponding
disconnected propagators 
do not enter into the quantities calculated in sections~\ref{NLOMass} and 
\ref{NLODecayConstant}, we do not write them explicitly here.

It is convenient to take $m_0^2\to \infty$ at this point
to decouple the $\eta'_I$. In the case of most current interest, with three staggered
flavors and the fourth root of the determinant taken to eliminate the extraneous
taste degree of freedom, $m^2_{\eta'_I} \cong m^2_0$ for large $m^2_0$ \cite{Aubin:2003mg}.
\Equation{DiscPropPrime} then becomes
\begin{equation}\label{eq:DiscProp}
        \[XY\]_{\rm disc}(p^2)=
-\frac{ 1}{3}\  \frac{(p^2 + m_{U_I}^2)(p^2 + m_{D_I}^2)(p^2 + m_{S_I}^2)}
                { (p^2 + m_{X}^2)(p^2 + m_{Y}^2) (p^2 + m_{\pi^0_I}^2)(p^2 + m_{\eta_I}^2) }
\ .
\end{equation}
Other disconnected valence-valence propagators are found from 
\eq{DiscProp} by substitution: Let $Y\to X$ for 
$\[XX\]_{\rm disc}$, $X\to Y$ for $\[YY\]_{\rm disc}$. If $m_u=m_d$, as in the MILC simulations
\cite{Aubin:2004fs,Aubin:2004wf}, then \begin{eqnarray}\label{eq:eigenvalues}
	m_{\pi^0_I}^2 &=& m_{U_I}^2= m_{D_I}^2\ ,\\
        m_{\eta_I}^2 & = & \frac{m_{U_I}^2}{3}+
        \frac{2m_{S_I}^2}{3} \ .\nn 
\end{eqnarray}

If we take the fourth root but keep the number $N_f$ of flavors arbitrary, then 
$m^2_{\eta'_I} \cong N_f m^2_0/3$, and the more general
version of \eq{DiscProp} is
\begin{equation}\label{eq:DiscPropGeneral}
        \[XY\]_{\rm disc}(p^2)=
-\frac{ 1}{N_f}\  \frac{\prod_{L=1}^{N_f}(p^2 + m_{L_I}^2)}
                { (p^2 + m_{X}^2)(p^2 + m_{Y}^2)\prod_{L'=1}^{N_f-1}  (p^2 + m_{L'_I}^2)}
\ ,
\end{equation}
where $L$ runs over diagonal flavor neutral states ($U$, $D$, $S$, \dots), and 
$L'$ run over the neutral mass eigenstates ($\pi_0$, $\eta$, \dots), excluding the $\eta'$.
%
\subsection{NLO valence-valence mass}
\label{NLOMass}
\begin{figure}[tbh]
\resizebox{5.0in}{!}{\includegraphics{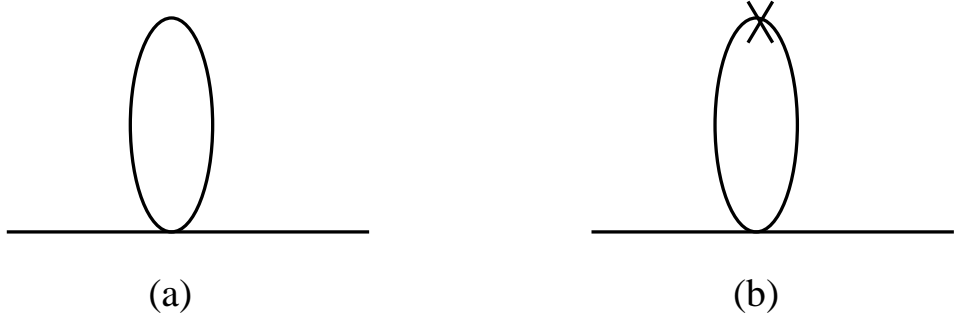}}
\caption{
Meson graphs for the $P^+$ self energy.  Graph (a) has a connected internal
propagator (\protect{\figref{ConnProp}} in the quark flow picture); while
graph (b) has a disconnected internal propagator
(\protect{\figref{DiscProp}}).
\label{fig:MesonGraphs}}
\end{figure}

%
We are interested in computing the one-loop correction to the meson made from
valence quark $x$ and $\bar y$, \ie a $P^+$.
The $P^+$ self energy comes from the meson graphs in \figref{MesonGraphs}. 
We thus need the four-meson vertices generated by the kinetic energy,
by the mass term, and by $\cU_V$, with at least one $P^+$ and one $P^-$
field.  

Expanding the kinetic energy, and including the minus sign for a vertex, 
gives the following terms with derivatives acting on both
$P^+$ and $P^-$: 
\begin{eqnarray}
\eqn{VKE1}
V^{(1)}_{\rm KE}&= \frac{\displaystyle 1}{\displaystyle 3f^2}\;\partial_\mu P^+\partial_\mu P^-\Big[
&\!\!\!2P^+P^- +X^2+Y^2 + R^\dagger_{\tilde xx}R_{\tilde xx} +
R^\dagger_{\tilde yx}R_{\tilde yx} + R^\dagger_{\tilde xy}R_{\tilde xy} +
R^\dagger_{\tilde yy}R_{\tilde yy} \nn \\
&&+ \sum_F\left(
Q^\dagger_{Fx}Q_{Fx} +
Q^\dagger_{Fy}Q_{Fy} \right) -2XY\Big]\ .
\end{eqnarray}
The fields in this expression are defined in \eq{PhiDef}.
The summation index $F$ runs over sea quarks, typically $u$, $d$, and $s$.  There is
also an implied sum over the taste index in the  product $Q^\dagger Q$.
As usual, terms with a derivative on $P^+$ or $P^-$, but not both, are not needed since the
corresponding diagrams in \figref{MesonGraphs} will vanish by symmetric 
momentum integration.
The kinetic energy terms with no derivatives on the $P^+$, $P^-$ fields are
\begin{eqnarray}
\eqn{VKE2}
V^{(2)}_{\rm KE}&= \frac{\displaystyle 1}{\displaystyle 3f^2}\;P^+P^-\;\Big[
&\!\!\! (\partial_\mu X)^2+(\partial_\mu Y)^2 + \partial_\mu R^\dagger_{\tilde xx}\,\partial_\mu R_{\tilde xx} +
\partial_\mu R^\dagger_{\tilde yx}\,\partial_\mu R_{\tilde yx} + 
\partial_\mu R^\dagger_{\tilde xy}\,\partial_\mu R_{\tilde xy}\\
&&\hspace{-0.7cm}+\partial_\mu R^\dagger_{\tilde yy}\,\partial_\mu R_{\tilde yy} 
+ \sum_F\left(
\partial_\mu Q^\dagger_{Fx}\,\partial_\mu Q_{Fx} +
\partial_\mu Q^\dagger_{Fy}\,\partial_\mu Q_{Fy} \right) -2\partial_\mu X\,\partial_\mu Y\Big].\nn
\end{eqnarray}
Similarly, the mass term and the ``mixed potential'' $\cU_V$ give, respectively,
\begin{eqnarray}
\eqn{VM}
V_{\rm mass}&= \frac{\displaystyle B}{\displaystyle 3f^2}\;P^+P^-\;\Bigg[
&\!\!\!(m_x+m_y)P^+P^- +(3m_x+m_y)X^2+(m_x+3m_y)Y^2 \\
&&\hspace{-2.2cm}+(3m_x+m_y)R^\dagger_{\tilde xx}R_{\tilde xx} 
+ 2\left(m_x+m_y\right)\left(R^\dagger_{\tilde yx}R_{\tilde yx} + R^\dagger_{\tilde xy}R_{\tilde xy} \right)
+ (m_x+3m_y)R^\dagger_{\tilde yy}R_{\tilde yy} \nonumber \\
&&\hspace{-1.0cm}+ \sum_F\left\{\left(2m_x+m_y+m_F\right)
Q^\dagger_{Fx}Q_{Fx} +
\left(m_x+2m_y+m_F\right)Q^\dagger_{Fy}Q_{Fy} \right\} \nonumber \\
&&\hspace{0.6cm} +2(m_x+m_y)XY\Bigg]\nn
\end{eqnarray}
and
\begin{equation}
\eqn{VUV}
V_{\rm Mix}= \frac{a^2\Delta_{\rm Mix}}{3f^2}\;P^+P^-\;
\sum_F\left(
Q^\dagger_{Fx}Q_{Fx} +
Q^\dagger_{Fy}Q_{Fx} \right) \ ,
\end{equation}
where we have used \eq{DeltaMix}.

\begin{figure}[tbh]
\resizebox{5.0in}{!}{\includegraphics{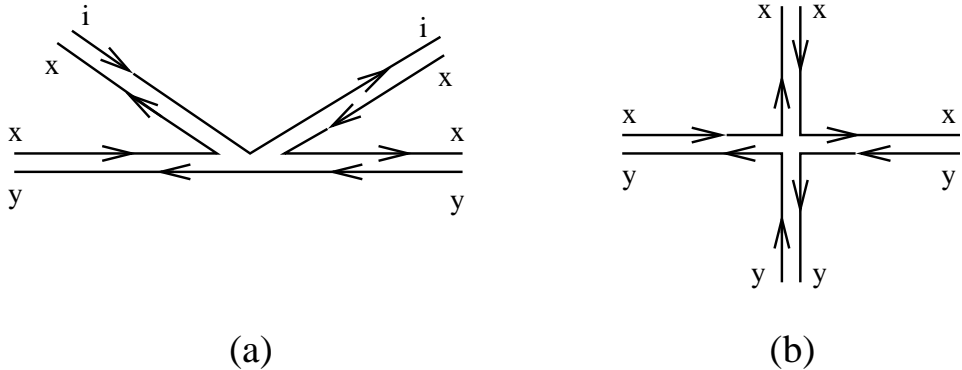}}
\caption{
Quark flow graphs corresponding to the four meson vertices in the
the $P^+$ self energy, \eqsfour{VKE1}{VKE2}{VM}{VUV}.  The horizontal $x$, $y$ lines
produce the external $P^{\pm}$ fields. Graph (a) represents terms where the
$P^+$ and $P^-$ are next to each other in the supertrace; an almost identical graph
with $x\leftrightarrow y$ is not shown. The free index $i$ represents any quark type, but the
numerical coefficient of the graph may depend on $i$.
Graph (b) represents terms where $P^+$ and $P^-$
are not next to each other in the supertrace.
\label{fig:Vertices}}
\end{figure}

By including the
ghost quark contributions we have guaranteed that quark loop terms from the valence quarks will
be canceled automatically, thereby accomplishing the partial quenching.  However, we still must
understand the meson diagrams at the quark-flow level \cite{Sharpe:1992ft} 
in order to adjust for the effects of
the fourth root procedure on the staggered quarks.  
If
we assume three flavors of sea quarks for definiteness, the fields $P^+$ and 
$P^-$ correspond to $\Phi_{45}$ and $\Phi_{54}$, respectively (cf.\ \eq{PhiDef}).  
The vertices, \eqsfour{VKE1}{VKE2}{VM}{VUV}, then come generically from two types of terms
in the supertrace of four $\Phi$ fields: terms where $P^+$ and $P^-$ are next to each other: 
$\sim \sum_i \Phi_{45} \Phi_{54} \Phi_{4i} \Phi_{i4}$ 
(or $\sim \sum_i \Phi_{54} \Phi_{45} \Phi_{5i} \Phi_{i5}$),
and terms where $P^+$ and $P^-$ are separated: $\sim \Phi_{45} \Phi_{55} \Phi_{54} \Phi_{44}$.
These can be represented by the vertices (a) and (b) in \figref{Vertices} respectively.
The index $i$ in \figref{Vertices}(a) should be
summed over all valence, ghost valence and sea quarks. The last contributions (proportional to $XY$) in
\eqsthree{VKE1}{VKE2}{VM} correspond to vertex \figref{Vertices}(b); while all other terms come
from the sum over $i$ in \figref{Vertices}(a).  

Note that the graphs in \figref{Vertices} represent quark flow only; the numerical
value of each graph depends on the term in the Lagrangian that generates it, and
may also depend on the free index $i$.  In particular the mixed potential $\cU_V$ generates
only vertex \figref{Vertices}(a) terms, and only gives non-zero coefficients of
terms where $i$ is a sea quark flavor.  This follows from
the fact that
$\tau_3$ in \eq{LsquareAdd} is proportional to the identity in the pure valence sector, 
and therefore $\cU_V$ reduces to a field-independent constant in that sector.

\begin{figure}[tbh]
\resizebox{5.0in}{!}{\includegraphics{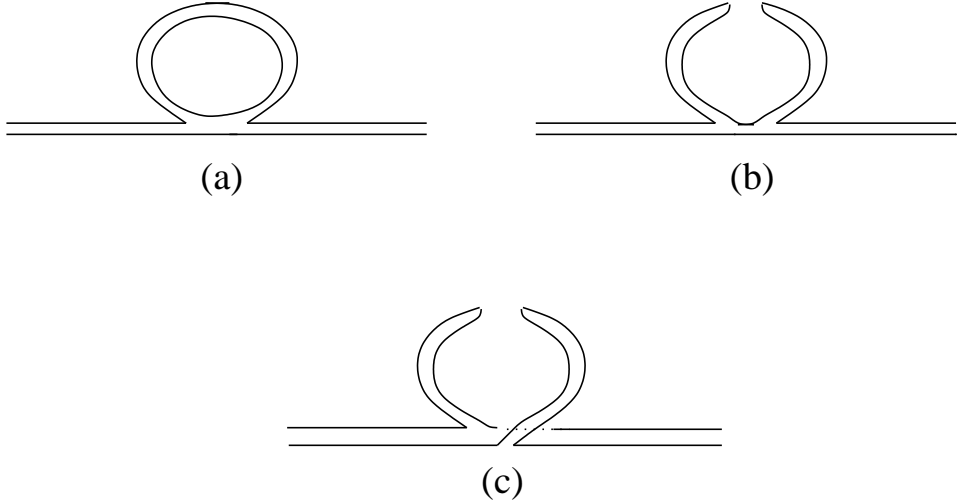}}
\caption{
Quark flow graphs corresponding to 
the $P^+$ self energy, \protect{\figref{MesonGraphs}}. Graphs (a) and (b) come from
connected and disconnected contractions, respectively,  of the internal meson lines in vertex
\protect{\figref{Vertices}}(a);  graph (c),  from the disconnected contraction in vertex
\protect{\figref{Vertices}}(b).  Iterations of sea quark loops in the disconnected
propagators, as in \protect{\figref{DiscProp}}, is implied in graphs (b) and (c).
\label{fig:Flow}}
\end{figure}

When two meson lines at the vertices are contracted, as in \figref{MesonGraphs}, the quark flow diagrams in \figref{Flow} result.  The connected contraction of \figref{Vertices}(a) gives
\figref{Flow}(a), which clearly involves an internal sea quark loop.  This means that terms
from \figref{Vertices}(a) where $i$ is a valence or ghost valence quark must cancel in the connected
contractions.  This arises algebraically from \eqsthree{VKE1}{VKE2}{VM} using \eq{ConnProp}:  Contractions
of $R^\dagger R$ terms cancel connected contractions of $XX$ and $YY$ and $P^+P^-$ terms.
This leaves just connected $Q^\dagger Q$ contractions (\figref{Flow}(a)), disconnected $XX$ and $YY$
contractions (\figref{Flow}(b)), and  disconnected $XY$ contractions (\figref{Flow}(c)).  

At this point it is easy to make the ``by-hand'' adjustment necessary to correspond with the
fourth-root procedure.  The only explicit sea quark loop in \figref{Flow} is in diagram (a),
so we just divide those terms by 4.  We already know such terms come only from
$Q^\dagger Q$ contractions; dividing them by 4 is equivalent to ignoring the implicit sum over
the 4 tastes in the $Q^\dagger Q$ terms.  There are also sea quark loop insertions in the
disconnected meson propagators in \figref{Flow}(b) and (c).  Such insertions have already
been corrected for the fourth root in \eq{DiscProp} or \eq{DiscPropGeneral}.

Let $\Sigma_{P^+}$ be the $P^+$ self energy, defined to be the negative of the sum of
self energy diagrams.  At NLO, we have
\begin{equation}
\eqn{Sigma}
\Sigma_{P^+}(p^2) = \frac{1}{16\pi^2 f^2}\left(\sigma_0\, m^2_{P} +\sigma_1\, p^2\right) +{\rm analytic\
 terms} \ ,
\end{equation}
where $\sigma_0$ and $\sigma_1$ come from the one-loop diagrams and are independent of momentum $p$.
Since we have not determined the NLO chiral Lagrangian, we cannot express the analytic contributions
in terms of chiral parameters.  For the quantities
of interest, however, it will be sufficient for our purposes to write down the most general contributions
consistent with the symmetries. 
Putting together the one-loop contributions from the vertices  \eqsfour{VKE1}{VKE2}{VM}{VUV}, we get
\begin{eqnarray}
\sigma_0 &=& -\frac{1}{3}\int \Bigg[\sum_F\left( \frac{1}{q^2+m^2_{Fx}}+ \frac{1}{q^2+m^2_{Fy}}\right)  \\
\eqn{sigma0}
&&\hspace{1.6truecm}+ \[XX\]_{\rm disc}\!(q^2)+\[YY\]_{\rm disc} \!(q^2)+4\[XY\]_{\rm disc}\! (q^2)\Bigg],\nn\\
\sigma_1 &=& -\frac{1}{3}\int \Bigg[\sum_F\left( \frac{1}{q^2+m^2_{Fx}}+ \frac{1}{q^2+m^2_{Fy}}\right)  \\
&&\hspace{1.6truecm}+ \[XX\]_{\rm disc}\!(q^2)+\[YY\]_{\rm disc} \!(q^2)-2\[XY\]_{\rm disc}\! (q^2)\Bigg],\nn
\eqn{sigma1}
\end{eqnarray}
where, as usual, $F$ runs over sea quark flavor (\eg $u$,$d$,$s$), and
\begin{equation}
\eqn{integral}
\int \equiv 16\pi^2\int \frac{d^4 q}{(2\pi)^4} \ ,
\end{equation}
suitably regulated. The disconnected propagator is given in \eq{DiscProp} or \eq{DiscPropGeneral}.
In \eq{sigma0}, we have dropped (constant) quartic divergences and
have used \eqsthree{VVmass}{VSmass}{SSmass} to replace factors of quark masses with the
corresponding meson masses. 
The simple identity 
\begin{equation}
\eqn{simple-identity}
\frac{q^2+m_X^2}{q^2+m_Y^2}\; \[XX\]_{\rm disc}\!(q^2)
=\frac{q^2+m_Y^2}{q^2+m_X^2}\;\[YY\]_{\rm disc} \!(q^2)=\[XY\]_{\rm disc}\! (q^2)
\end{equation}
has allowed us to remove all explicit factors of $q^2$ in the integrands involving disconnected propagators.

We now focus explicitly on computing the 
$P^\pm$ mass.
The chiral symmetry in the valence sector implies that the analytic contributions to
$(m_{P}^{\rm NLO})^2$
must be proportional to the tree-level $m_P^2\propto (m_x+m_y)$.  
In the continuum limit, these contributions
just go over to the standard form in terms of the Gasser-Leutwyler parameters 
$L_i$ \cite{Gasser:1984gg}. At finite lattice
spacing, the only possible new NLO analytic term 
is $\cC a^2m_P^2$, where $\cC$ is an unknown constant that depends on the details
of the lattice action in both the valence and the sea sectors.  Indeed it is easy to write down
terms that will appear in the NLO Lagrangian and will contribute to $\cC$,  for example,
\begin{eqnarray}
\eqn{someNLOterms}
& a^2 \left\langle P_V\partial_\mu \Sigma P_V\partial_\mu \Sigma^\dag\right\rangle
\ , \\
        & a^2 \cU_S \left\langle\partial_{\mu}\Sigma \partial_{\mu}\Sigma^{\dagger}\right\rangle \ ,\nn \\
 & a^2 \cU_S \left\langle \Sigma M^{\dagger} + M \Sigma^{\dagger}\right\rangle 
\nonumber   ,\nn
\end{eqnarray}
with $\cU_S$ given in \eq{US}.
The one-loop corrections to the $P^\pm$ mass squared are then found by evaluating
$\Sigma_{P^+}(p^2)$ at $p^2\!=\!- m_{P}^2$, giving
\begin{eqnarray}
\eqn{Pmass1}
\frac{(m_{P}^{\rm NLO})^2 }{B(m_x+m_y)}
& =&1+\frac{1}{16\pi^2 f^2}
\left(\sigma_0-\sigma_1\right) 
 + \frac{16B}{f^2}\left(2L_8-L_5\right)\left(m_x+m_y\right) \\
        &&\hspace{0.8truecm}
         +\frac{32B}{f^2}\left(2L_6-L_4\right)
         \sum_f m_f + a^2 \cC  ,\nn
\end{eqnarray}
where, from \eqstwo{sigma0}{sigma1},
\begin{equation}
\eqn{MassIntegral}
\sigma_0-\sigma_1 = -2\int \[XY\]_{\rm disc}\! (q^2) \ .
\end{equation}

The integral in \eq{MassIntegral} can be evaluated in terms of the chiral logarithm and residue
functions defined in Refs.~\cite{Aubin:2003mg,Aubin:2003uc}.  For completeness we include the
definitions here. The chiral logarithm functions, coming from integration over a single and double pole,
respectively, are:
\begin{eqnarray}\label{eq:chiral-log1}
        \ell( m^2)& \equiv & m^2 \left(\ln \frac{m^2}{\Lambda_\chi^2}
         + \delta_1(mL)\right)\ ,
        \\*
        \label{eq:chiral-log2}
         \tilde \ell(m^2)& \equiv & -\left(\ln
        \frac{m^2}{\Lambda_\chi^2} + 1\right) + \delta_3(mL) \ ,\nn
\end{eqnarray}
where $\Lambda_\chi$ is the chiral scale, $L$ is the spatial dimension, and
the finite volume correction terms are
\cite{Bernard:2001yj}
\begin{eqnarray}\label{eq:delta1}
        \delta_1(mL) & \equiv & 4
                \sum_{\vec r\ne 0}
                \frac{K_1(|\vec r|mL)}{mL|\vec r|} \ , \\*
        \label{eq:delta3}
        \delta_3(mL) & \equiv& 2 \sum_{\vec r\ne 0}
                K_0(|\vec r|mL)\ .\nn
\end{eqnarray}
$K_0$ and $K_1$ are Bessel functions of imaginary argument, and $\vec  r$,
which labels the various
periodic images, is a three-dimensional vector with integer components.

The residue functions $R^{[n,k]}_j$ allow one to write integrals over
ratios of products of $(q^2+m^2)$ terms, such as $\[XY\]_{\rm disc}\!(q^2)$, as integrals over
single poles. They are defined by
\begin{equation}\label{eq:residues}
        R_j^{[n,k]}\left(\left\{\cM\right\}\!;\!\left\{\mu\right\}\right)
         \equiv  \frac{\prod_{a=1}^k (\mu^2_a- m^2_j)}
        {\prod_{i=1}^{'n} (m^2_i - m^2_j)}\ .
\end{equation}
The residues are a function of two sets of masses, the ``denominator'' set
$\{\cM\}=\{m_1,m_2,\ldots, m_n\}$ and the ``numerator'' set $\{\mu\}=\{\mu_1,\mu_2,\ldots,\mu_k\}$.
The indices $j$ and $i$, $1\le j,i \le n$, refer to particular denominator masses; the
prime on the product in the denominator of \eq{residues} means that $i=j$ is omitted.
In cases of degeneracy, we also need the double-pole residue functions, $D^{[n,k]}_{j,\ell}$:
\begin{equation}\label{eq:residues2}
        D_{j, \ell}^{[n,k]}\left(\left\{\cM\right\}\!;
        \!\left\{\mu\right\}\right) \equiv -\frac{d}{d m^2_{\ell}}
        R_{j}^{[n,k]}\left(\left\{\cM\right\}\!;
        \!\left\{\mu\right\}\right)\ .
\end{equation}

We can now write out \eq{Pmass1} explicitly in various useful cases. In the $N_f=3$ partially quenched case,
with no mass degeneracies,
we have
\begin{eqnarray}\label{eq:mPQ111}
        \frac{(m^{\rm NLO}_{P^+})^2}
        {B\left( m_x+m_y \right)}&=& 1 +
        \frac{1}{16\pi^2f^2}\Bigg(
        \frac{2}{3} \sum_{j=1}^4 R^{[4,3]}_{j}(\{\cM^{[4]}_{XY,I}\};\{\mu^{[3]}_I\})\;
         \ell(m^2_{j}) \Bigg)+ \\
        &&\hspace{-1.2truecm}
 + \frac{16B}{f^2}\left(2L_8-L_5\right)\left(m_x+m_y\right) 
         +\frac{32B}{f^2}\left(2L_6-L_4\right)
         \left(m_u+m_d+m_s\right)+ a^2 \cC .\nn
\end{eqnarray}
The index $j$ is summed over the denominator masses, as it will also be in subsequent
cases;  
the mass-set arguments are 
\begin{eqnarray}\label{eq:mass-setsPQ111}
        \{\cM^{[4]}_{XY,I}\}& \equiv & \{ m_{X}, m_{Y}, m_{\pi^0_I}, m_{\eta_I} \} \ , \\
        \{ \mu^{[3]}_I \} &\equiv & \{ m_{U_I},m_{D_I},m_{S_I}   \} \ .\nn
\end{eqnarray}

\Equation{mPQ111} is identical to the corresponding 
continuum partially quenched result with no degeneracies \cite{Sharpe:2000bc}, 
except for the explicit discretization term $a^2\cC$ and the
fact that the neutral sea-sea mesons are specified to be taste singlets.
For us, ``no degeneracies'' means that none of the seven meson masses in \eq{mass-setsPQ111}
are equal. We have chosen the normalization of 
quark masses in the chiral Lagrangian, \eq{LOChiralLagrangian}, so that the same constant 
$B$ appears in the
relation between quark and meson masses in both sea and valences cases. However, 
degeneracies between sea-sea and valence-valence mesons in \eq{mass-setsPQ111} would {\em not}\ 
imply equal valence and sea quark masses, because of the splitting of the taste-singlet sea-sea mesons,
\eq{SSmass}.

Taking $m_u=m_d\equiv\hat m  \Rightarrow m_{U_I}=m_{D_I}=m_{\pi^0_I}$, but with no other degeneracies, then gives
\begin{eqnarray}\label{eq:mPQ21}
        \frac{(m^{\rm NLO}_{P^+})^2}
        {B\left( m_x+m_y \right)}&=& 1 +
        \frac{1}{16\pi^2f^2}\Bigg(
        \frac{2}{3} \sum_{j=1}^3 R^{[3,2]}_{j}(\{\cM^{[3]}_{XY,I}\};\{\mu^{[2]}_I\})\;
         \ell(m^2_{j}) \Bigg)+ \\
        &&\hspace{-1.2truecm}
 + \frac{16B}{f^2}\left(2L_8-L_5\right)\left(m_x+m_y\right) 
         +\frac{32B}{f^2}\left(2L_6-L_4\right)
         \left(2\hat m+m_s\right)+ a^2 \cC\ ,\nn
\end{eqnarray}
with 
\begin{eqnarray}\label{eq:mass-setsPQ21}
        \{\cM^{[3]}_{XY,I}\}& \equiv & \{ m_{X}, m_{Y}, m_{\eta_I} \}\ ,\\
        \{ \mu^{[2]}_I \} &\equiv & \{ m_{U_I},m_{S_I}   \} \ .\nn
\end{eqnarray}

When $m_u=m_d\equiv\hat m $ and 
$m_x=m_y \Rightarrow m_X=m_Y$, but no degeneracies between
sea-sea and valence-valence mesons, we have
a ``partially quenched pion'' with
\begin{eqnarray}\label{eq:mPQpion}
        \frac{(m^{\rm NLO}_{P^+})^2}
        {2Bm_x}&=& 1 +
        \frac{1}{16\pi^2f^2}\,
        \frac{2}{3}\, \Bigg(R^{[2,2]}_{1}(\{\cM^{[2]}_{X,I}\};\{\mu^{[2]}_I\})\,\tilde\ell(m^2_X) \\
&&\hspace{2.7truecm}+\sum_{j=1}^2 D^{[2,2]}_{j,1}(\{\cM^{[2]}_{X,I}\};\{\mu^{[2]}_I\})\,
         \ell(m^2_{j}) \Bigg)+ \nonumber \\
        &&\hspace{0.7truecm}
 + \frac{16B}{f^2}\left(2L_8-L_5\right)\left(2m_x\right) 
         +\frac{32B}{f^2}\left(2L_6-L_4\right)
         \left(2\hat m+m_s\right)+ a^2 \cC\ ,\nn
\end{eqnarray}
where
\begin{equation}\label{eq:mass-setsPQpion}
        \{\cM^{[2]}_{X,I}\} \equiv  \{ m_{X}, m_{\eta_I} \} \ ,
\end{equation}
and $\{ \mu^{[2]}_I \}$ is given by \eq{mass-setsPQ21}.

At finite lattice spacing, the cases that most resemble the full (unquenched) theory are ones
with degeneracies among the valence-valence and sea-sea mesons in \eq{mass-setsPQ111}.
For current purposes, we might call a 
``full pion'' one with $m_X=m_Y=m_{U_I}=m_{D_I}=m_{\pi^0_I}$, which requires
$m_x=m_y$ and $m_u=m_d=\hat m$, but $m_x>\hat m$, since the taste splitting $\Delta_{\pi^{0}_{I}}$ in \eq{SSmass} 
is positive.\footnote{This is at least true for the simulations carried out by the 
MILC collaboration.  } 
In this case, we have
\begin{eqnarray}\label{eq:mfullpion}
        \frac{(m^{\rm NLO}_{\pi^+})^2}
        {2Bm_x}&=& 1 +
        \frac{1}{16\pi^2f^2}\,
         \left(\ell\big(m^2_{\pi^0_I}\big) - \frac{1}{3}\,\ell\big(m^2_{\eta_I}\big) \right)+  \\
 &&+ \frac{16B}{f^2}\left(2L_8-L_5\right)\left(2m_x\right) 
         +\frac{32B}{f^2}\left(2L_6-L_4\right)
         \left(2\hat m+m_s\right)+ a^2 \cC\ ,\nn
\end{eqnarray}
where we have used \eq{eigenvalues} to simplify the residues.
Similarly, calling a ``full kaon'' a meson with $m_X=m_{U_I}=m_{D_I}$ and $m_Y=m_{S_I}$, results in
\begin{eqnarray}\label{eq:mfullkaon}
        \frac{(m^{\rm NLO}_{K})^2}
        {B(m_x+m_y)}&=& 1 +
        \frac{1}{16\pi^2f^2}\, \frac{2}{3}\,\ell\big(m^2_{\eta_I})+ \\
 &&+ \frac{16B}{f^2}\left(2L_8-L_5\right)\left(m_x+m_y\right) 
         +\frac{32B}{f^2}\left(2L_6-L_4\right)
         \left(2\hat m+m_s\right)+ a^2 \cC\ .\nn
\end{eqnarray}
\Equations{mfullpion}{mfullkaon} clearly approach the standard results \cite{Gasser:1984gg} as $a\to0$.

\subsection{NLO valence-valence pseudoscalar decay constant}
\label{NLODecayConstant}
The decay constant $f_P$ of the $P^\pm$ meson  is defined by the matrix element of the
corresponding axial current, $j_{\mu5}^{P}$, 
\begin{equation}\label{eq:matrix_element}
        \left\langle 0 \left| j_{\mu5}^{P}
        \right| P(p) \right\rangle =-i
        f_{P} p_{\mu} \ .
\end{equation}
In the LO chiral theory of \eq{LOChiralLagrangian}, the axial current is given by\footnote{We use the Noether current corresponding to axial vector rotations as our partially conserved axial vector current.  This is justified if the analogous current is used in the underlying lattice theory. A convenient  method to 
construct this current in numerical simulations is described in Ref.\ \cite{Hasenfratz:2002rp}. 
Alternatively, one may employ the corresponding covariant pseudoscalar density
to define the decay constant \cite{Hasenfratz:2002rp}. 
On the other hand, using a conserved but not covariant axial vector current 
would result in extra terms proportional to $am_{x}$ and $am_{y}$, which are not captured in the following results.}
\begin{eqnarray}\label{eq:current}
        j_{\mu5}^{P} &= &\frac{-i f^2}{4}
                \left\langle\lambda \left(
                \partial_{\mu}\Sigma\Sigma^{\dagger} +
                \Sigma^{\dagger}\partial_{\mu}\Sigma
                \right)\right\rangle\ .
\end{eqnarray}
Here $\lambda$ projects out the 
appropriate flavors:
With three sea-quark flavors as in \eq{PhiDef}, the valence quarks $x$ and $y$ correspond
to indices $4$ and $5$  of $\Phi$, and
then  $\lambda_{ij} = \delta_{i5}\delta_{j4}$.

At NLO, the decay constant has the form
\begin{equation}\label{eq:fpi1}
        \frac{f^{\rm NLO}_{P}}{f} = 1 + \frac{1}{16\pi^2 f^2} \delta\!
        f_{P} +{\rm analytic\ terms}  \ .
\end{equation}
The term $\delta\!f_{P}$ comes from the one-loop diagrams, and the analytic contributions are generated
at tree-level by NLO terms in the chiral Lagrangian and corresponding corrections to the current
in \eq{current}.
There are two contributions to $\delta\! f_{P}$,
\begin{equation}
\delta\! f_{P} = \delta\! f^{\rm current}_{P} + \frac{1}{2}\delta\! Z_P \ ,
\end{equation}
where $\delta\! f^{\rm current}_{P}$ comes from diagrams generated directly by 
expanding the current in 
\eq{current} to cubic order in $\Phi$, and 
$\delta\! Z_{P}$ is the one-loop wave function renormalization.
From \eq{Sigma} we have
\begin{equation}\label{eq:wave_ren_contr}
        \delta\! Z_P
        \equiv -16\pi^2f^2\;
               \frac{d\Sigma_{P^+}(p^2)}{dp^2} = -\sigma_1  \ .
\end{equation}
As in Ref.~\cite{Aubin:2003uc}, a straightforward calculation shows that
$\delta\! f^{\rm current}_{P} = -2\,\delta\!  Z_P$. 
From \eq{sigma1}, we then have
\begin{eqnarray}\label{eq:deltafpi}
        \delta\! f_{P} = 
         \frac{3}{2}\sigma_1 
&=& -\frac{1}{2}\int \Bigg[\sum_F\left( \frac{1}{q^2+m^2_{Fx}}+ \frac{1}{q^2+m^2_{Fy}}\right)\\
&&\hspace{1.6truecm}+ \[XX\]_{\rm disc}\!(q^2)+\[YY\]_{\rm disc} \!(q^2)-2\[XY\]_{\rm disc}\! (q^2)\Bigg] \ ,\nn
\end{eqnarray}
with $F$ summed over sea quark flavors.

The analytic terms in \eq{fpi1} come only from NLO terms in the chiral Lagrangian with derivatives,
which affect the decay constant either directly, through wave function renormalization,
or indirectly, by leading to higher corrections to the axial current.  Thus, $\mcal O(a^4)$ corrections
to the chiral Lagrangian have no effect on \eq{fpi1}. There will however be
analytic terms from $\mcal O(p^4)$, $\mcal O(mp^2)$, and $\mcal O(a^2p^2)$ Lagrangian corrections. The former
two are identical to those in the continuum, and produce terms proportional to
quark masses.  The effects of the latter corrections on $f_P$ can be absorbed into a single term 
proportional to $a^2$.  We thus have,
\begin{equation}\label{eq:fpi}
        \frac{f^{\rm NLO}_{P}}{f} =  1 + \frac{1}{16\pi^2 f^2} \delta\!  f_{P} + 
        \frac{8B}{f^2}L_5\left( m_x + m_y \right)
        + \frac{16B}{f^2}L_4\sum_F m_F +a^2\cF \ , 
\end{equation}
where  $L_4$ and $L_5$ are standard \cite{Gasser:1984gg}, $\cF$ is a new constant,
and $\delta\!  f_{P}$ is given by \eq{deltafpi}. Because Lagrangian terms
of $\cO(ma^2)$ do not affect $\cF$,  
it is easy to see from the discussion surrounding
\eq{someNLOterms} 
that $\cF$ is independent of the corresponding
constant $\cC$ occurring in the expression for the meson mass, \eq{Pmass1}.
Like $\cC$, $\cF$ depends on the details of the lattice action in both the
sea and valence sectors.

We can now write out the NLO expression for the decay
constant in various special cases. 
In the $N_f=3$ partially quenched case, with no mass degeneracies,
we have
\begin{eqnarray}
\label{eq:fPQ111}
        \frac{f^{\rm NLO}_{P}}{f} & = &  1
        + \frac{1}{16\pi^2 f^2}
         \Biggl[-\frac{1}{2}\sum_{F} \left[\ell\left(m^2_{Fx}\right)+\ell\left(m^2_{Fy}\right)\right] \\
        &&\hspace{1.4truecm}+ \frac{1}{6}\Biggl(R^{[3,3]}_{1}(\{\cM^{[3]}_{X,I}\};\{\mu^{(3)}_{I}\})
        \tilde\ell(m^2_{X})
        +R^{[3,3]}_{1}(\{\cM^{[3]}_{Y,I}\};\{\mu^{(3)}_{I}\})
        \tilde\ell(m^2_{Y})
        \nonumber \\* 
&&\hspace{1.4truecm} +\sum_{j=1}^3
        D^{[3,3]}_{j,1}(\{\cM^{[3]}_{X,I}\};\{\mu^{(3)}_{I}\})\ell(m^2_{j}) 
+\sum_{j=1}^3D^{[3,3]}_{j,1}(\{\cM^{[3]}_{Y_I}\};\{\mu^{(3)}_{I}\})\ell(m^2_{j}) \nonumber \\*
       &&\hspace{1.4truecm} -2\sum_{j=1}^4R^{[4,3]}_{j}(\{\cM^{[4]}_{XY,I}\};\{\mu^{(3)}_{I}\})\ell(m^2_{j})
         \Biggr)        \Biggr] \nonumber \\* 
        &&\hspace{.7truecm} + \frac{8B}{f^2}L_5\left( m_x + m_y \right)
        + \frac{16B}{f^2}L_4\sum_Fm_F + a^2\cF
         \ ,\nn
\end{eqnarray}
where $F$ runs over $u$, $d$, and $s$. The mass sets 
$\{\cM^{[4]}_{XY,I}\}$ and $\{\mu^{(3)}_{I}\}$ are given by \eq{mass-setsPQ111}, and
\begin{eqnarray}\label{eq:mass-setsX3Y3}
        \{\cM^{[3]}_{X,I}\}& \equiv & \{ m_{X}, m_{\pi^0_I}, m_{\eta_I} \}\ ,\\
        \{\cM^{[3]}_{Y,I}\}& \equiv & \{ m_{Y}, m_{\pi^0_I}, m_{\eta_I} 
\} \ .\nn 
\end{eqnarray}

With $m_u=m_d\equiv\hat m  \Rightarrow m_{U_I}=m_{D_I}=m_{\pi^0_I}$, but no other degeneracies, 
the result is
\begin{eqnarray}\label{eq:fPQ21}
        \frac{f^{\rm NLO}_{P}}{f} & = &  1
        + \frac{1}{16\pi^2 f^2}
         \Biggl[-\frac{1}{2}\left[2\ell\left(m^2_{ux}\right)+\ell\left(m^2_{sx}\right)
         +2\ell\left(m^2_{uy}\right)+\ell\left(m^2_{sy}\right)\right]\\
        &&\hspace{1.4truecm}+ \frac{1}{6}\Biggl(R^{[2,2]}_{1}(\{\cM^{[2]}_{X,I}\};\{\mu^{(2)}_{I}\})
        \tilde\ell(m^2_{X})
        +R^{[2,2]}_{1}(\{\cM^{[2]}_{Y,I}\};\{\mu^{(2)}_{I}\})
        \tilde\ell(m^2_{Y})
        \nonumber \\*
&&\hspace{1.4truecm} +\sum_{j=1}^2
        D^{[2,2]}_{j,1}(\{\cM^{[2]}_{X,I}\};\{\mu^{(2)}_{I}\})\ell(m^2_{j})
+\sum_{j=1}^2D^{[2,2]}_{j,1}(\{\cM^{[2]}_{Y_I}\};\{\mu^{(2)}_{I}\})\ell(m^2_{j}) \nonumber \\*
       &&\hspace{1.4truecm} -2\sum_{j=1}^3R^{[3,2]}_{j}(\{\cM^{[3]}_{XY,I}\};\{\mu^{(2)}_{I}\})\ell(m^2_{j})         \Biggr)        \Biggr] \nonumber \\*
        &&\hspace{.7truecm} + \frac{8B}{f^2}L_5\left( m_x + m_y \right)
        + \frac{16B}{f^2}L_4\left( 2\hat m + m_s\right) + a^2\cF
         \ .\nn
\end{eqnarray}
Here, $\{\cM^{[3]}_{XY,I}\}$ and $\{\mu^{(2)}_{I}\}$ are given in \eq{mass-setsPQ21}; while
$\{\cM^{[2]}_{X,I}\}$ is defined in \eq{mass-setsPQpion} (for $\{\cM^{[2]}_{Y,I}\}$ take $X\to Y$).

For a ``partially quenched pion'' with $m_x=m_y$, there is considerable simplification
because the disconnected contributions in \eq{deltafpi} will cancel.
Taking in addition $m_u=m_d\equiv \hat m$ for simplicity, we have
\begin{eqnarray}\label{eq:fPQpion}
        \frac{f^{\rm NLO}_{P}}{f} & = & 1
        + \frac{1}{16\pi^2 f^2}
         \Big[-2\ell\left(m^2_{ux}\right)-\ell\left(m^2_{sx}\right)\Big]
\\
        &&\hspace{.7truecm} + \frac{8B}{f^2}L_5\left( 2m_x \right)
        + \frac{16B}{f^2}L_4\left( 2\hat m + m_s\right) + a^2\cF
        \nn \ .
\end{eqnarray}
There is no obviously preferred way here to define a ``full pion'' to make the
NLO corrections take on a continuum-like form.  In the $a\to0$ limit, the splitting
$a^2\Delta_{\rm Mix}$ in \eq{VSmass} will vanish, and the logarithm terms will clearly approach
the standard form \cite{Gasser:1984gg}: $-2\ell\left(m^2_{\pi}\right)-\ell\left(m^2_{K}\right)$.
At finite lattice spacing, we can choose $m_x$ so that $m^2_{ux}$ and $m^2_{sx}$ have the
masses of the sea-sea pion and sea-sea kaon of any one taste, but there seems to be
no advantage in doing that.  In particular, the value of $m_x$ so chosen will {\em not}\/ be the
same in general as the value needed to give the logarithms in the meson mass their continuum-like
form, \eq{mfullpion}.

For the kaon, it makes some sense to define a ``full kaon'' as we did in Sec.~\ref{NLOMass}:
$m_X=m_{U_I}=m_{D_I}=m_{\pi^0_I}$ and $m_Y=m_{S_I}$.  This at least gives the disconnected contributions
the form they would have in the continuum.  We then have
\begin{eqnarray}\label{eq:ffullkaon}
        \frac{f^{\rm NLO}_{K}}{f} & = &  1
        + \frac{1}{16\pi^2 f^2}
         \Biggl[-\ell\left(m^2_{ux}\right)-\frac{1}{2}\ell\left(m^2_{sx}\right)
         -\ell\left(m^2_{uy}\right)-\frac{1}{2}\ell\left(m^2_{sy}\right) \\
        &&\hspace{2.9truecm}+ \frac{1}{4}\ell\left(m^2_{\pi^0_I}\right)
 +\frac{1}{2}\ell\left(m^2_{S_I}\right)
-\frac{3}{4}\ell\left(m^2_{\eta_I}\right)
                      \Biggr] \nonumber \\*
        &&\hspace{.7truecm} + \frac{8B}{f^2}L_5\left( m_x + m_y \right)
        + \frac{16B}{f^2}L_4\left( 2\hat m + m_s\right) + a^2\cF
         \ .\nn
\end{eqnarray}
In the continuum limit, $m_{ux}=m_{\pi^0_I}\equiv m_{\pi}$,
 $m_{sx}=m_{uy}\equiv m_{K}$, $m_{\eta_I} \equiv m_{\eta}$,
and
 $m_{sy}=m_{S_I}$, thereby reproducing the known result \cite{Gasser:1984gg}.

\section{Discussion}
\label{summary}

Our results for 2+1 sea quark flavors are currently the most relevant ones, since they can 
be applied to simulations using the existing configurations generated by the MILC collaboration. 
Eqs.\ \pref{eq:mPQpion} and \pref{eq:fPQpion} describe the quark mass 
and lattice spacing $a$ dependence of the pion 
masses and decay constants, and these expressions can be directly fitted to lattice data 
obtained with Ginsparg-Wilson valence fermions.

The number of unknown fit parameters in these expressions is fairly small. For instance, the pion mass depends on the usual low-energy constants of continuum \chpt\ ($f$, $B$ and two familiar combinations of Gasser-Leutwyler coefficients), the sea-sea meson masses $m^{2}_{\pi^{0}_{I}}, m^{2}_{\eta_{I}}$, the sea quark masses $\hat{m}, m_{s}$ and the constant $\cC$. At one loop order we can express the sea quark mass combination $2\hat{m}+ m_{s}$ through leading order sea-sea meson masses (cf.\ eq.\ \pref{eq:SSmass}): 
\bea
B( 2\hat{m}+ m_{s} )& = & \frac{1}{2} m^{2}_{\pi^{+}_{5}} + m^{2}_{K_{5}^{+}}\,.
\eea
These masses as well as $m^{2}_{\pi^{0}_{I}}$ and  $m^{2}_{\eta_{I}}$ have already been measured 
by the MILC collaboration and are therefore not unknown parameters.\footnote{The measurement of the singlet meson masses is difficult because disconnected diagrams contribute to the correlator. For $m_u=m_d$, however,  there are no disconnected contributions to the $\pi^0_{I}$
propagator, and its mass is degenerate with the  $\pi^+_{I}$ mass. The $\eta_{I}$ mass is also not affected by disconnected diagrams if $m_u=m_d \not= m_s$ and the limit $m_0 \to \infty$ is taken. In that case it is consistent to employ eq.\ \pref{eq:eigenvalues} where the $m_{S_I}$ and $m_{U_I}$ masses are from connected diagrams only. In reality $m_0^2$ is not infinity, and there can be
$\eta - \eta'$ mixing, which  would be proportional to $(m_s -
m_u)/m_0^2$.  Such corrections are not taken into account in the MILC determination of $m_{\eta_{I}}$.} 
The only true unknown parameter in addition to the ones from continuum \chpt\ is thus the constant $\cC$.

Similar statements apply to the pion decay constant in eq.\ \pref{eq:fPQpion}. 
Even though the masses $m_{ux}^{2}, m_{sx}^{2}$ 
of the valence-sea mesons have not been measured yet, they can be straightforwardly determined from the propagator of the mixed meson and a linear fit to the leading order mass formula in \eq{VSmass}. Using this information leaves one additional parameter, the constant $\cF$, besides the familiar parameters of continuum \chpt.

Let us compare our results for Ginsparg-Wilson valence quarks with the corresponding expressions for staggered valence quarks. The one loop expression for the Goldstone pion $\pi_{5}^{+}$ in the 2+1 flavor case reads (see eq.\ (75) in Ref. \cite{Aubin:2003mg})  
\begin{eqnarray}\label{eq:pi_K_final_deg}
	\frac{(m^{\rm NLO}_{\pi^+_5})^2}{2B\hat{m}}
	&=&1 + \frac{1}{16\pi^2 f^2}\Biggl(\ell(m_{\pi^0_I}^2)
	 - \frac{1}{3}\ell(m_{\eta_I}^2) \Biggr) \\*
 &&+ \frac{16B}{f^2}\left(2L_8-L_5\right)\left(2\hat{m} \right)
	+\frac{32B}{f^2}\left(2L_6-L_4\right)
	 \left(2\hat{m}+m_s \right) + a^2 \tilde{\cC}\nonumber\\*
	 && -\frac{1}{16\pi^2 f^2}\Biggl(4\ell(m^2_{\pi^0_V})
	+ \frac{2a^2\delta'_V}{m_{\eta'_V}^2 - m^2_{\eta_V}}\biggl[
	 \frac{m_{S_V}^2 - m^2_{\eta_V}}
	{m_{\pi^0_V}^2 -
	m^2_{\eta_V}} \ell(m^2_{\eta_V}) 
	-  \frac{m_{S_V}^2 - m^2_{\eta'_V}}
	{m_{\pi^0_V}^2 -m^2_{\eta'_V}}  \ell(m^2_{\eta'_V})
	\biggr]\nonumber\\*
	&&\hspace{1.2cm}+4\ell(m^2_{\pi^0_A})
	+ \frac{2a^2\delta'_A}{m_{\eta'_A}^2 - m^2_{\eta_A}}\biggl[
	 \frac{m_{S_A}^2 - m^2_{\eta_A}}
	{m_{\pi^0_A}^2 -
	m^2_{\eta_A}} \ell(m^2_{\eta_A}) 
	-  \frac{m_{S_A}^2 - m^2_{\eta'_A}}
	{m_{\pi^0_A}^2 -m^2_{\eta'_A}}  \ell(m^2_{\eta'_A})
	\biggr]\Biggr)\ .\nn
\end{eqnarray}
The first two rows of this expression give the corresponding result in eq.~\pref{eq:mfullpion} for Ginsparg-Wilson valence quarks. The remaining contributions involve many more sea-sea meson masses as well as the two ``hairpin'' parameters $\delta'_V$ and $\delta'_A.$\footnote{These two parameters are combinations of the low-energy constants in the potential ${\cal U}^{\prime}$: $\delta'_{V} = C_{2V}-C_{5V}$, and analogously for $\delta'_{A}$ \cite{Aubin:2003mg}.} These parameters cannot be expressed in terms of leading order 
charged meson masses and are therefore true unconstrained fit parameters. 

The one-loop result for the pion decay constant -- given in eq.\ (27) in Ref.\ \cite{Aubin:2003uc} -- also has contributions proportional to $\delta'_V$ and $\delta'_A$:
\begin{eqnarray}\label{eq:final_21_pion_result}
	\frac{f^{\rm NLO}_{\pi^+_5}}{f} && =  1 + 
	\frac{1}{16\pi^2 f^2}
	 \sum_{b}\frac{
	  -2\ell(m^2_{\pi^0_b})- \ell(m^2_{K^+_b})}{16}
	 + \frac{16B}{f^2}\left( 2\hat{m} + m_s\right)L_4
	+ \frac{16B}{f^2}\hat{m} L_5
	+a^2 \tilde{\cF} \\*
&&
	\!\!\!+ \frac{1}{16\pi^2 f^2}
	 \Biggl(2a^2\delta'_V\Biggl[  \frac{m^2_{S_V}-m^2_{\eta_V}}
	{(m^2_{\pi^0_V}-m^2_{\eta_V})(m^2_{\eta'_V}-m^2_{\eta_V})}
	\ell(m^2_{\eta_V})
	+ \frac{m^2_{S_V}-m^2_{\eta'_V}}
	{(m^2_{\pi^0_V}-m^2_{\eta'_V})(m^2_{\eta_V}-m^2_{\eta'_V})}
	\ell(m^2_{\eta'_V})\nonumber \\* && 
	\hspace{1.4cm}+  \frac{m^2_{S_V}-m^2_{\pi^0_V}}
	{(m^2_{\eta_V}-m^2_{\pi^0_V})(m^2_{\eta'_V}-m^2_{\pi^0_V})}
	\ell(m^2_{\pi^0_V}) \Biggr]\nn\\*&&
	\!\!\!+ 2a^2\delta'_A\Biggl[  \frac{m^2_{S_A}-m^2_{\eta_A}}
	{(m^2_{\pi^0_A}-m^2_{\eta_A})(m^2_{\eta'_A}-m^2_{\eta_A})}
	\ell(m^2_{\eta_A})
	+ \frac{m^2_{S_A}-m^2_{\eta'_A}}
	{(m^2_{\pi^0_A}-m^2_{\eta'_A})(m^2_{\eta_A}-m^2_{\eta'_A})}
	\ell(m^2_{\eta'_A})\nonumber \\* && 
	\hspace{1.4cm}+  \frac{m^2_{S_A}-m^2_{\pi^0_A}}
	{(m^2_{\eta_A}-m^2_{\pi^0_A})(m^2_{\eta'_A}-m^2_{\pi^0_A})}
	\ell(m^2_{\pi^0_A}) \Biggr]\Bigg) \ .\nn
\end{eqnarray}
The factor of $1/16$ in the first line is canceled in the continuum limit by the
sum over $b$, which runs over all sixteen different meson tastes. 
As was the case for the pseudoscalar masses, 
the corresponding expression for the mixed theory, eq.\ \pref{eq:fPQpion}, is much simpler and does not involve the contributions proportional to the hairpin parameters $\delta'_V$ and $\delta'_A$. 
Note that the constants $\tilde{\cC},\tilde{\cF}$ in eqs. \pref{eq:pi_K_final_deg} and \pref{eq:final_21_pion_result} are different from $\cC$ and $\cF$ in eqs.\ (\ref{eq:mfullpion}) and \pref{eq:fPQpion}.

Obviously the functional dependence in the expressions for staggered valence quarks is much more complicated and involves more undetermined parameters than in the corresponding results for the mixed theory. However, the physical low-energy constants, the Gasser-Leutwyler coefficients, enter the expressions in the same way. Therefore, as already pointed out in Ref.\ \cite{Bar:2002nr}, mixed simulations may be used to extract these phenomenologically relevant parameters from numerical lattice simulations. 

The NLO formulas computed in Secs.~\ref{NLOMass} and \ref{NLODecayConstant} also make
concrete a rather obvious fact about mixed action theories: 
At non-zero lattice spacing there is no way to define equality of valence and sea quark masses in order to have all properties that might be desired of a ``full'' (unquenched) theory.  The lattice theory will
always have some features of partial quenching, and it is only in the continuum
limit that the full theory is obtained.  
Depending on the purpose one may wish to choose various definitions to match the sea and valence quark masses. Since the scalar correlator is very sensitive to the effects of partial quenching \cite{Bardeen:2001jm}, it has been proposed to choose the valence quark mass such that the scalar correlator does not have a negative contribution \cite{Bowler:2004hs}. For the results reported in Ref.\ \cite{Renner:2004ck} the masses were chosen so that the valence-valence pion mass coincides with the Goldstone pion mass made of staggered quarks, \ie $m^{2}_{\pi^{+}}= m^{2}_{\pi^{+}_{5}}$. The NLO results have suggested yet another definition with 
$m^{2}_{\pi^{+}}= m^{2}_{\pi^{+}_{I}}$. In this case, as we have seen in Sec.~\ref{NLOMass}, 
some chiral logarithms resemble their continuum form and one might expect smaller partial quenching effects than with other definitions.\footnote{With the definition in Ref.\ \cite{Renner:2004ck} the scalar correlator becomes negative  \cite{Schroers:privateC}, a clear signal for partial quenching.} From a theoretical point of view all these definitions are equally good since they guarantee that full unquenched QCD is reached in the continuum limit. Practically, they differ with respect to the size of the partial quenching effects at non-zero lattice spacing, and the quark mass tuning can be rather difficult to achieve, depending on, for example, the statistical errors in the observables used for the matching.

Nevertheless, at least in the context of chiral perturbation theory, there is no fundamental difficulty with using mixed action theories to simulate QCD. The effects of finite lattice spacing can be controlled by first fitting to the chiral forms of the type derived here, and then extrapolating to the continuum limit.
Furthermore, as in the pure staggered case \cite{Bernard:2004ab}, we expect
the chiral and continuum limits to commute in the mixed theory 
for any quantity that has a well-defined chiral limit in the continuum.

\begin{acknowledgments}
We thank M.~Golterman and S.~Sharpe for helpful discussions. O.\ B.\ and C.~B. gratefully acknowledge support of the Benasque Center of Science, where part of this work was done during the workshop ``Matching light quarks to hadrons''.
This work is supported in part by the Grants-in-Aid for
Scientific Research from the Japanese Ministry of Education, 
Culture, Sports, Science and Technology 
(Nos. 13135204, 15204015, 15540251, 16028201), the University of Tsukuba Research Project and by the US Department
of Energy under grant numbers DE-FG02-91ER40628, W-7405-ENG-36, 
DE-FG03-96ER40956/A006 and DE-FG02-91ER40676.

\end{acknowledgments}

\appendix
\section{Construction of mixed four-fermion operators} 
\label{appendix1} 
In this appendix we construct the mixed four-fermion operators in eq.\ \pref{MixedQuartic} that enter the Symanzik effective theory at $\mcal O(a^{2}$). We closely follow the procedure and notation in Ref.\ \cite{Lee:1999zx} where the four-fermion operators 
made of sea quarks only were constructed. The method determines first all lattice four-fermion operators without derivatives and mass insertions that are singlets under the symmetries of the lattice theory. Taking the continuum limit of these terms results in the allowed continuum operators that appear in the Symanzik effective action.

First we convert the staggered fields into hypercube fields, since these fields yield the proper continuum fields when the lattice spacing is sent to zero. Following Refs.~\cite{Gliozzi:1982ib,Duncan:1982xe,Kluberg-Stern:1983dg} the lattice is divided into hypercubes containing 16 sites whose coordinates are written as\footnote{We use lattice units and set $a=1$ in this appendix.} 
\bea
x_{\mu} &=&2y_{\mu} +\eta_{\mu}.
\eea
The hypercube vector $\eta$ labels the sites within the hypercube and its components $\eta_{\mu}$ are either 0 or 1. In terms of the site variables $\chi$, $\overline{\chi}$ and the gauge links $U$ the hypercube fields are defined by 
\bea
\psi_{S,\alpha a}(y) &=&\frac{1}{2}\sum_{\eta}\Gamma_{\eta}^{\alpha a}U(2y,2y+\eta)\chi_{S}(2 y +\eta),\\
\overline{\psi}_{S,\alpha a}(y) &=& \frac{1}{2}\sum_{\eta}\overline{\chi}_{S}(2 y +\eta) U^{\dagger}(2y,2y+\eta)\Gamma_{\eta}^{*\alpha a}, \nn
\eea
where $U(2y,2y+\eta)$ denotes a product of link variables along a path going from $2y$ to $2y+\eta$, and
\bea
\Gamma_{\eta} &=& \gamma_{1}^{\eta_{1}} \gamma_{2}^{\eta_{2}} \gamma_{3}^{\eta_{3}} \gamma_{4}^{\eta_{4}}\,.
\eea
The indices  $\alpha$ and $a$ represent the Dirac and taste index, respectively (we suppress the flavor and the color index). 

Using the hypercube fields we now construct all mixed four-fermion operators $O_{\rm 4f}$ 
that are allowed by the symmetries and that correspond to dimension 6 operators in the continuum limit. 
Four-fermion operators that contain derivatives and/or quark masses are higher than $\mcal O(a^2)$
in the Symanzik action and can be ignored here.
Since no quark mass appears the 
operators must be invariant under the full chiral symmetries of the massless lattice theory.

The construction proceeds in five steps \cite{Lee:1999zx}:
\begin{enumerate}
\item Multiply a sea quark bilinear on a hypercube by a valence bilinear at the same lattice point and sum over all hypercubes,
\bea\label{GeneralFF}
O_{\rm 4f}(\Gamma_{S},\Gamma_{V}) &=& \sum_{y} \Big(\overline{\psi}_{S}(y)\Gamma_{S}\,\psi_{S}(y)\Big) \Big(\overline{\psi}_{V}(y)\Gamma_{V}\,\psi_{V}(y) \Big)\,.
\eea
The flavor symmetries dictate that the sea quark bilinear is an $SU(N_{f})$ singlet while the valence bilinear is a singlet under  $SU(N_{V}|N_{V})$. $\Gamma_{S}$ represents an arbitrary tensor product $\gamma_{A}\otimes t^{a}\otimes\xi_{\alpha}$ of a gamma matrix acting in Dirac space, a color generator $t^{a}$ and an $SU(4)$ taste group generator $\xi_{\alpha}$. Similarly, $\Gamma_{V}$ denotes an arbitrary combination  $\gamma_{B}\otimes t^{b}$ acting on the valence fields. It does not include a slot for a taste matrix since the valence fields do not have the taste degree of freedom.

In eq.\ \pref{GeneralFF} all the sea quark indices are contracted with $\Gamma_{S}$ while all the valence indices are contracted with $\Gamma_{V}$, so the operator truly is a product of two bilinears. One can write down other operators that do not have this simple structure. For example, one could contract the color indices of $\psi_{S}$ and $\overline{\psi}_{V}$ and the indices of  $\psi_{V}$ and $\overline{\psi}_{S}$. Similarly one can contract the Dirac indices in such a ``twisted'' manner. However, all these 
operators are redundant \cite{Bar:2003mh}. Making use of Fierz identities one can always ``untwist'' these operators and bring them into the form in eq.\ \pref{GeneralFF}. 

On each hypercube there are $16^{2}$ possibilities to form a valence field bilinear. The one chosen
in eq.\ \pref{GeneralFF} involves only the fields at the lattice point $y$ where the staggered
hypercube field lives.  Close to the continuum, all other valence bilinears
can be written as the one in eq.\ \pref{GeneralFF}, plus terms involving
derivatives, which we can drop.

\item The sum over all hypercubes in eq.\ \pref{GeneralFF} makes the operator $O_{\rm 4f}$ invariant under lattice translations by one hypercube, \ie $y\rightarrow y + 1$. 
In order to obtain the part that is invariant under single site translations we apply the projection operator
\bea\label{DefTranlation}
{\cal P} & = & \prod_{\mu} \frac{1}{2}\left(1 + T_{S}^{\mu}\otimes T_{V}^{\mu}\right) ,
\eea
with $T_{S}^{\mu}\otimes T_{V}^{\mu}$ being the translation operator in the $\mu$-direction. The translation operator acts differently in the sea and valence sector. In the valence sector it is a trivial shift of the fields,
\bea
T_{V}^{\mu} \psi_{V}(x) & =& \psi_{V}(x+\hat{\mu}),\\
T_{V}^{\mu} \overline{\psi}_{V}(x)&=& \overline{\psi}_{V}(x+\hat{\mu}),\nn
\eea
where $\hat{\mu}$ denotes the unit vector in the $\mu$ direction.
In the sea sector it involves transformations of the spin and taste degrees of freedom. Explicitly \cite{Jolicoeur:1986ek} 
\bea
T_{S}^{\mu} \psi_{S}(y) & =& \frac{1}{2} \Big[ (I\otimes \xi_{\mu} - \gamma_{{\mu5}}\otimes\xi_{5})  \psi_{S}(y)
+  (I\otimes \xi_{\mu} + \gamma_{{\mu5}}\otimes\xi_{5})  \psi_{S}(y + 2\mu)\Big],\label{SeaShift1}\\[1ex]
T_{S}^{\mu} \overline{\psi}_{S}(y) & =& \frac{1}{2} \Big[ \overline{\psi}_{S}(y) (I\otimes \xi_{\mu} + \gamma_{{\mu5}}\otimes\xi_{5})  
+  \overline{\psi}_{S}(y + 2\mu)(I\otimes \xi_{\mu} - \gamma_{{\mu5}}\otimes\xi_{5}) \Big].\label{SeaShift2}\nn
\eea
Note that the translation operators for different directions commute when acting
on staggered bilinears, so the order of them is irrelevant in the product in eq.~\pref{DefTranlation}.

Applying the projection operator eq.~\pref{DefTranlation} results in many terms with derivatives, which we can neglect. Acting with $T_{V}^{\mu}$ on the valence bilinear gives
\bea
\overline{\psi}_{V}(y)\Gamma_{V}\,\psi_{V}(y) &\rightarrow& \Big(\overline{\psi}_{V}(y)+ \nabla_{\mu}^{f}\overline{\psi}_{V}(y)\Big) \Gamma_{V}\,\Big(\psi_{V}(y) +\nabla_{\mu}^{f}\psi_{V}(y)\Big), 
\eea
where we have introduced the usual nearest-neighbor forward difference operator in $\mu$ direction, $\nabla_{\mu}^{f}$. Similarly, acting with $T_{S}^{\mu}$ on sea quark bilinear produces many derivative terms. Using eqn.\ \pref{SeaShift1} and \pref{SeaShift2} one straightforwardly establishes 
\bea
\hspace{-.5truecm}\sum_{y}\overline{\psi}_{S}(y)\Gamma_{S}\,\psi_{S}(y) &\rightarrow& \sum_{y}\overline{\psi}_{S}(y) 
( I\otimes \xi_{\mu})\Gamma_{S}( 
I\otimes \xi_{\mu})\psi_{S}(y) + \mbox{derivative terms}. 
\eea 
Using these two results we find
\bea
{\cal P} \Big[O_{\rm 4f}(\Gamma_{S},\Gamma_{V})\Big] &=& O_{\rm 4f}(\tilde{\Gamma}_{S},\Gamma_{V})+ \mbox{derivative terms}.
\eea
The matrix $\tilde{\Gamma}_{S}$ differs from ${\Gamma}_{S}$ only in the taste matrix: $\xi_{\alpha}$ is replaced by the average
\bea
\tilde{\xi}_{\alpha} = \frac{1}{16} \sum_{i=1}^{16} \xi_{i}^{\dagger}\xi_{\alpha}\xi_{i},
\eea
where the sum runs over all sixteen elements of the Clifford algebra. Only the identity $\xi_{\alpha}=I$ survives this average; for the other 15 taste matrices the average is zero. Thus we conclude that we only need to consider sea quark bilinears in eq.\ \pref{GeneralFF} that are taste singlets.

\item Next we impose the constraint that the operators must be color singlets. There are only two ways to form such singlets. Either the color group generators in $\Gamma_{S}$ and $\Gamma_{V}$ are equal and a summation over the generator index is performed, or the identity matrix is inserted instead. 

\item Now we form linear combinations of $O_{\rm 4f}(\Gamma_{S},\Gamma_{V})$ that are singlets under  the hypercubic symmetry group of the lattice ($\pi/2$ rotations and reflections) and also charge conjugation. The transformation properties of the staggered fields in the hypercube notation are listed in Refs.\ \cite{Verstegen:1985kt,Luo:1996vt}. Since the matrix $\Gamma_{S}$ is trivial in taste space these transformations act in spin space only and their form is the same as for continuum Dirac spinors. We therefore find that the gamma matrices in $\Gamma_{S}$ and $\Gamma_{V}$ must be equal with their open indices being properly contracted in order to form scalars under rotations and reflections. 

\item Finally we select the operators that are invariant under the chiral symmetries. Each bilinear must be separately invariant under the full chiral symmetry group. This excludes all gamma matrices but the vector and axial vector.

\end{enumerate}
This procedure produces all mixed four-fermion operators without derivatives that are singlets under all lattice symmetries. Taking the continuum limit one finds the four invariant operators listed in eq.\ \pref{MixedQuartic}.

\appendix

%

\end{document}